\documentclass[jap,reprint]{revtex4-2}

\usepackage{graphicx} 
\usepackage{dcolumn}   
\usepackage{bm}        
\usepackage{amssymb}   
\usepackage{mathtools}
\usepackage{textcomp}	
\usepackage{natbib}      
\usepackage{amsmath}
\usepackage{color}

\begin{document}
\bibliographystyle{apsrev4-2}

\def\mytitle{Temperature dependence of the optical phonon reflection band in GaP}
\title{\mytitle}
\author{Nuwanjula S.\ Samarasingha}\email[Email:]{sandu87@nmsu.edu}
\homepage[URL: ]{http://ellipsometry.nmsu.edu}
\author{Stefan Zollner} \email[Email:]{zollner@nmsu.edu}
\affiliation{Department of Physics, New Mexico State University, P.O.\ Box 30001, Las Cruces, NM 88003, USA}
\date{\today}

\begin{abstract}
We explore the effect of temperatures between 80 and 720~K on the energy and linewidth of zone-center transverse (TO) and longitudinal (LO) optical phonons in bulk gallium phosphide (GaP) using Fourier transform infrared ellipsometry from 0.03 to 0.60 eV. We extract the optical phonon parameters of GaP by fitting the ellipsometric angles with the Lowndes-Gervais model, which applies two different broadening parameters to the TO and LO phonons. In GaP, the two-phonon density of states is larger for the decay of TO phonons than for LO phonons. Therefore, we observed a larger TO phonon broadening (compared to the LO phonon) and an asymmetric reststrahlen line shape. This would lead to a negative imaginary part of the dielectric function just above the LO phonon energy, but the addition of two-phonon absorption avoids this. We find a temperature dependent redshift and broadening of TO and LO phonons with increasing temperature due to thermal expansion and anharmonic phonon-phonon scattering, involving three and four phonon decay processes. We also investigate the temperature-dependence of the high-frequency dielectric constant. Its variation is explained by thermal expansion and the temperature dependence of the Penn gap.
\end{abstract}


\maketitle 


\section{Introduction}

Most insulators and non-conducting polar semiconductors exhibit bands of high reflectance in the far- or mid-infrared spectral range. These so-called {\em reststrahlen} bands extend from the energy of transverse optical (TO) phonons to that of the corresponding longitudinal optical (LO) phonons. They have been studied extensively for many materials.\cite{Lo70,Sc04,BeUn68,LoYu05,GePi74,Ba64,Ba68,ZoPa19} 

One open question regarding these restrahlen bands is the relative magnitude of the broadening parameters of the TO and LO phonons. Lowndes\cite{Lo70} and Schubert\cite{Sc04} explain that the sum of the broadenings $\gamma_{\rm TO}$ of the TO phonons must be smaller than the sum of the broadenings $\gamma_{\rm LO}$ of the LO phonons. Berreman and Unterwald\cite{BeUn68} call this the ``passivity'' condition, ensuring that the imaginary part of the dielectric function $\epsilon\left(\omega\right)$ does not become negative. This condition is satisfied for alkali halides.\cite{Lo70} 

It is often violated for other materials, however, as shown by Lockwood, Yu, and Rowell\cite{LoYu05} for six different zinc blende semiconductors using infrared reflectance measurements at oblique incidence combined with a derivative analysis technique, where $\gamma_{\rm TO}$$>$$\gamma_{\rm LO}$ is found for AlAs, GaP, InP, InAs, and InSb. For highly ordered materials, such as zinc blende semiconductors, these broadenings are lifetime broadenings, related to the decay of optical phonons into acoustic ones.\cite{De98} For GaP, the decay of TO phonons into two acoustic phonons is very fast, because a maximum of the two-phonon density of states (DOS) occurs at the TO energy.\cite{De00,VeLo01} By contrast, the two-phonon DOS is lower at the LO energy and therefore $\gamma_{\rm TO}$$>$$\gamma_{\rm LO}$ for GaP. As we will show below, the resulting negative $\epsilon_2$ above the LO energy is compensated by two-phonon absorption.

To further investigate the broadenings of the reststrahlen bands in semiconductors, we measured the temperature dependence of the dielectric function of GaP in the mid-infrared spectral region from 80 to 720~K. We selected GaP for our study, because its reststrahlen band is at rather high energies (due to the low mass of the P atom) and Fourier-transform infrared (FTIR) ellipsometry measurements can therefore be performed using commercial instrumentation. Furthermore, the reststrahlen band is not distorted by two-phonon absorption between the TO and LO energies.\cite{TrCa15} We find a redshift and broadening of the TO and LO phonons due thermal expansion and anharmonic phonon decay. 

Gallium phosphide (GaP) is an indirect III/V compound semiconductor with a band gap $E_g$ of approximately 2.25~eV at room temperature.\cite{RiPi19,ApBr14} It crystallizes in the zinc blende structure\cite{BoHa79,RiPi19} with the cubic space group $T_d^2$. GaP has two atoms per primitive cell, resulting in three-fold degenerate acoustic and optical phonon modes at $\Gamma$. The polar Fr\"ohlich interaction splits the optical modes into a TO doublet and an LO singlet,\cite{Sn00,Bi62} which are both Raman- and infrared-active. This thermally stable indirect wide band gap material is an excellent semiconductor for optoelectronic and photonic applications, especially in light-emitting diodes (LEDs),\cite{SoNi98} detectors, solar cells, and high-temperature transistors.\cite{ApBr14} It is therefore important to study the optical properties of this semiconductor, including the effects of cryogenic and elevated temperatures. 

The vibrational properties of bulk GaP have already been studied theoretically and experimentally by several authors. Using FTIR reflectance, Lockwood {\em et al.}\cite{LoYu05} investigated TO and LO phonons in several III-V semiconductor materials, including bulk GaP in the reststrahlen region at room temperature. Bairamov {\em et al.}\cite{BaKi74,BaPa79} measured the effect of temperature on the optical phonon frequency and linewidth with Raman scattering and explained their results with the anharmonic decay of optical phonons into two or three acoustic phonons. They found a redshift and increasing broadening with increasing temperature from 4.2 K to 550 K and also an increasing asymmetry of the TO phonon at high temperatures. Debernardi\cite{De98,De00} calculated the temperature and pressure dependence of the Raman linewidths and energies of the TO and LO phonons from 0 to 325 K. From the observation of first order Raman spectra, Mooradian and Wright\cite{MoWr66} found almost no phonon frequency shift between room temperature and helium temperature. The spectral range of most of these works is around the reststrahlen band (250-550 cm$^{-1}$) of bulk GaP. However, several other theoretical and experimental studies of optical phonons in bulk GaP exist,\cite{KaCh76,AbGl89,Ha67,MaPa91,GiJa76} including inelastic neutron scattering data and shell model calculations.\cite{BoHa79} Some of them identified not only the long wavelength optical phonons at the $\Gamma$ point of the Brillouin zone, but also zone edge phonons.\cite{Po83,Ba68,PoIz83}

\section{Experimental methods and models}
\label{methods}

A bulk undoped single side polished GaP wafer grown by the liquid encapsulated Czochralski method with 0.5~mm thickness, (111) surface orientation, and a (110) flat was obtained commercially (MTI Corporation, Richmond, CA). The sample had n-type conductivity with an electron concentration on the order of $n$=5$\times$10$^{16}$~cm$^{-3}$ and a resistivity of about 0.3~$\Omega$cm. By comparison, the electron concentrations of GaP used for similar studies in the literature range from 10$^{16}$ to 10$^{17}$~cm$^{-3}$. The electron density may influence the observed phonon line broadenings. 

The ellipsometric angles $\Psi$ and $\Delta$ of the as-received GaP wafer were acquired (without cleaning) on a J.\ A.\ Woollam FTIR Mark II ellipsometer from 0.03-0.60 eV (mid and near-infrared spectral regions) with 1 cm$^{-1}$  resolution from 80-720 K with 25 K step size (30 measurements) inside an ultra-high vacuum cryostat (Janis ST-400) with diamond windows (Diamond Materials GmbH, Freiburg, Germany) at 70$^\circ$ angle of incidence.\cite{Sc04} At each temperature, measurements were performed with 15 positions per revolution of the rotating compensator. To increase the signal-to-noise ratio, we averaged 50 interferometer mirror scans at each compensator position. Systematic errors were reduced with $P$$=$$\pm{45}^\circ$ polarizer angles and also zone averaging the analyzer ($A$$=$$0^\circ,90^\circ$). WVASE32 (J.\ A.\ Woollam Co., Lincoln, NE) and IGOR Pro (WaveMetrics, Lake Oswego, OR) scientific data analysis software were used to analyze our data.

We first measured $\Psi$ and $\Delta$ in air at three different incidence angles of 60$^\circ$, 65$^\circ$, and  70$^\circ$ to calibrate the cryostat windows.\cite{Ab20} The GaP sample was then attached to a copper cold finger using metal clamps. The temperature was measured with two type-E (nickel-chromium/constantan) thermocouples. One located near the cryogen reservoir was used to control the temperature with a Lakeshore temperature controller. The second thermocouple was directly attached to the surface of the GaP sample and measured the temperature of the sample surface. The difference between both thermocouple readings was up to 60 K at the highest temperatures. 

\begin{figure}
\includegraphics[width=\columnwidth]{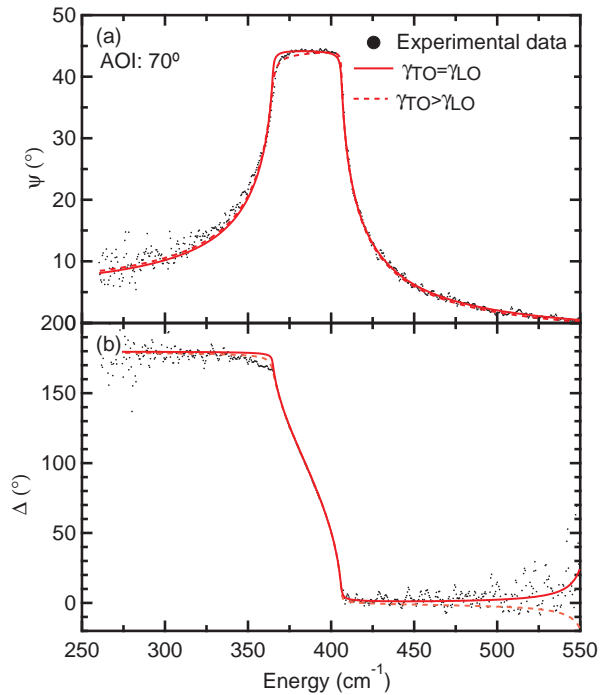}
\caption{Ellipsometric angles $\psi$ and $\Delta$ at 300~K in air at 70$^\circ$ angle of incidence (AOI) for bulk GaP with a surface layer. The symbols show experimental data. Solid and dashed lines show the result of a fit using Eq.\ (\ref{Lowndes}).}
\label{elipsoangle}
\end{figure}

The cryostat was then sealed and pumped to a pressure below 10$^{-5}$ Torr. 
In order to reduce the surface contamination, the sample was heated to 700 K for several hours with a 50 $\Omega$ resistor installed in the cryostat. This anneal may change the surface reconstruction, but no serious degradation of the sample is expected.\cite{SuLa03} The cryostat was then allowed to cool down to room temperature while continuing to pump. Once a sufficiently low base pressure of 10$^{-9}$-10$^{-8}$ Torr was reached, we started to cool the system down to 80~K using liquid nitrogen. Measurements were then taken from low to high temperatures in steps of about 25~K. Typical results at 300~K in air are shown in Fig.\ \ref{elipsoangle}. 

The static and high-frequency dielectric constants $\epsilon_s$ and $\epsilon_\infty$ and the optical phonon parameters (amplitude $A$, TO and LO phonon energies $\omega_{\rm TO}$ and $\omega_{\rm LO}$, and corresponding broadenings $\gamma_{\rm TO}$ and $\gamma_{\rm LO}$) were obtained by fitting the ellipsometric angles of GaP at each temperature either with a single Lorentzian or with the Lowndes-Gervais model\cite{Lo70,GePi74}
\begin{equation}
\epsilon(\omega)=\epsilon_{\infty}\frac{\omega_{LO}^{2}-\omega^2-i\gamma_{LO}\omega}{\omega_{TO}^{2}-\omega^2-i\gamma_{TO}\omega}.
\label{Lowndes}
\end{equation}
Details of this approach are given in Ref.\ \citenum{ZoPa19} and in the supplementary material. 
A thin surface overlayer with variable thickness (about 35 \AA), described as a 50/50 mixture of GaP and voids within the Bruggeman effective medium approximation, was also included in the model.\cite{ToHi16,Fu07}

\begin{figure}
\includegraphics[width=\columnwidth]{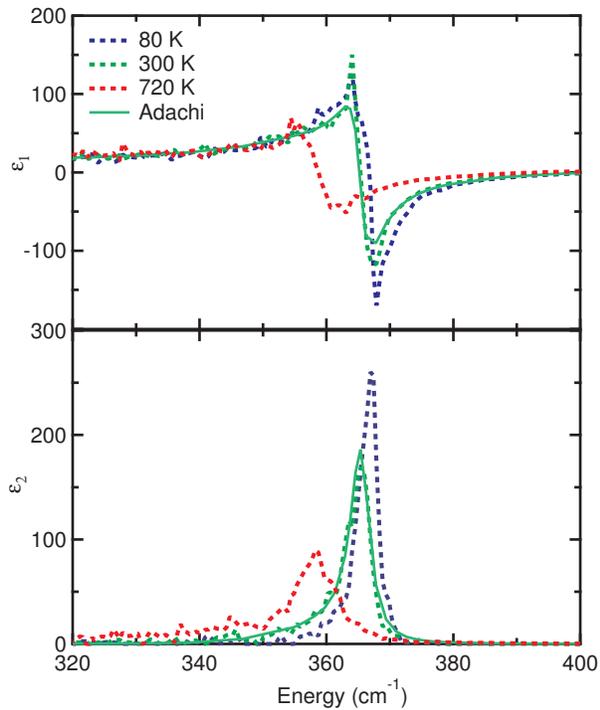}
\caption{Real and imaginary parts of the complex dielectric function of bulk GaP at three different temperatures versus photon energy (dashed), determined from a ``point-by-point'' fit, in comparison with room temperature results tabulated by Adachi\cite{Ad13} (solid).}
\label{opticalconstant}
\end{figure}

Once the surface layer thickness and phonon parameters have been determined from this model for each temperature, it is also possible to fix the surface layer thickness and fit the ellipsometric angles with the optical constants at each photon energy as parameters. This so-called ``point-by-point fit'' leads to results in Fig.\ \ref{opticalconstant}. It can be seen clearly that the extrema of $\epsilon_1$ and $\epsilon_2$ show a redshift, increasing broadening, and decreasing amplitude with increasing temperature. Tabulated optical constants\cite{Ad13} are shown for comparison. 

\section{RESULTS AND DISCUSSION}

\subsection{Optical phonon parameters at 300 K}

The Lorentz model (\ref{Lorentz}) allowed us to find $\epsilon_\infty$, $A$, $\omega_{\rm TO}$, and $\gamma_{\rm TO}$ at room temperature, either by fitting the ellipsometric angles with the experimental errors (model 1) or only the imaginary part $\left<\epsilon_2\right>$ of the pseudo-dielectric function (model 2). Both models yield similar results, see Table \ref{para_Table}, but $\gamma_{\rm TO}$ is smaller when fitting the ellipsometric angles (model 1).  
The Lorentzian broadening affects both the lower and upper corner of the reststrahlen band. If $\gamma_{\rm TO}$$>$$\gamma_{\rm LO}$, model (1) will find an intermediate Lorentzian broadening close to the average of $\gamma_{\rm TO}$ and $\gamma_{\rm LO}$. On the other hand, if we only fit $\left<\epsilon_2\right>$ in model (2), we find the larger broadening $\gamma_{\rm TO}$ of the TO phonon. This explains why models (1) and (2) report different values for $\gamma_{\rm TO}$, see Table \ref{para_Table}.

These models (1-2) assume equal broadenings for the TO and LO phonons. Results from the Lorentz model (1) are shown by the solid lines in Fig.\ \ref{elipsoangle}. This model (1) fits the ellipsometric angles near the LO energy well, but not near the TO energy. The Lorentzian line shape is symmetric, while the ellipsometric angle $\psi$ is more rounded at the TO energy (indicating a larger broadening) than at the LO energy. 

We also determined the energy and broadening of the LO phonon by fitting the loss function (\ref{Loss}) with a Lorentzian, see Fig.\ \ref{loss}. Results are shown in model (3) of Table \ref{para_Table}. We find that $\gamma_{\rm TO}$ is two to four times larger than $\gamma_{\rm LO}$. This motivates the use of the Lowndes-Gervais model (4), see Eq.\ (\ref{Lowndes}), to fit the ellipsometric angles, because it treats $\gamma_{\rm TO}$ and $\gamma_{\rm LO}$ as two independent parameters. These results are shown by the dashed line in Fig.\ \ref{elipsoangle} and in model (4) of Table \ref{para_Table}. 

\begin{table}
\caption{Phonon parameters at 300~K determined with different models: (1) TO phonon parameters obtained from the Lorentz model by fitting the ellipsometric angles with the experimental errors. $\omega_{\rm LO}$ was calculated from Eq.\ (\ref{LOTO}). (2) Same, but from fitting only $\left<\epsilon_2\right>$. (3) LO phonon parameters from a Lorentzian fit to the pseudo-loss function. (4) Results from the Lowndes-Gervais model, Eq.\ (\ref{Lowndes}). Data from the literature is also shown. Probable errors are listed in parentheses.}
\label{para_Table}
\begin{ruledtabular}
\begin{tabular}{lllllll}
Model & $\epsilon$$_\infty$ & A & $\omega_{\rm TO}$ & $\gamma_{\rm TO}$  & $\omega_{\rm LO}$  & $\gamma_{\rm LO}$ \\
& & & (cm$^{-1}$) & (cm$^{-1}$) & (cm$^{-1}$) & (cm$^{-1}$) \\
\hline
(1)  & 9.02(1) & 1.950(2) & 364.81(3)& 1.61(3) & 402.33(6) & NA\\
(2)  & NA & 1.97(3) & 365.24(3)& 3.86(8) & NA & NA\\
(3) & NA & 0.0211(1) & NA& NA & 402.3(2) & 0.88(1)\\
(4) &  9.02(1) & NA & 364.81(3)& 2.91(4) & 402.33(2) &0.90(3)\\
\hline
Ref.\ \citenum{LoYu05} & 9.24(2) & & 366.3(3) & 2.6(6) & 402.50(5) & 1.2(1) \\
Ref.\ \citenum{Ba68} & 9.09 & 1.92 & 365.3 & 1.1 & 402.2 \\
Ref.\ \citenum{BaKi74}\footnotemark[1]\footnotemark[2] & & & 364.5 & 3.5 & 402.5 & 2 \\
Ref.\ \citenum{MoWr66}\footnotemark[1] & & & 367.3 & & 403.0\\
Ref.\ \citenum{PaCo71} & 9.091 & & 363.4 \\
Ref.\ \citenum{KuBr84}\footnotemark[3] & & & & $>$4 & & 0.76 \\
\end{tabular}
\end{ruledtabular}
\footnotetext[1]{Raman scattering.}
\footnotetext[2]{$n$=9$\times$10$^{16}$~cm$^{-3}$ (increased LO broadening).}
\footnotetext[3]{Coherent anti-Stokes Raman scattering.}
\end{table}

Our value of $\epsilon_\infty$=9.02 is only about 0.8\% smaller than $\epsilon_\infty$=9.091 determined by Parsons and Coleman.\cite{PaCo71,BoGu98} A better accuracy should not be expected for spectroscopic ellipsometry and can only be obtained with minimum-deviation prims measurements.\cite{Bo65} Using the LST relation (\ref{LST}), we find a static dielectric constant $\epsilon_s$=10.97, also slightly (1.6\%) smaller than $\epsilon_\infty$=11.147 found with far-infrared measurements.\cite{PaCo71,BoGu98} The temperature dependence of these dielectric constants will be discussed later within the context of Fig.\ \ref{epssepinf}. Our value for $\omega_{\rm LO}$ agrees to within 0.1 cm$^{-1}$ with Lockwood {\em et al.}\cite{LoYu05} and Palik,\cite{PaCo71,BoGu98} but differences in the literature for $\omega_{\rm TO}$ are several wave numbers. This is unexpected, since the upper end of the reststrahlen band near $\omega_{\rm LO}$ depends on the angle of incidence and on the free electron concentration (see supplementary material), while the lower end of this band should unambiguously yield $\omega_{\rm TO}$. 

It can be seen in Fig.\ \ref{elipsoangle} that the reststrahlen band for bulk GaP is asymmetric, a clear indication that the TO and LO phonon broadenings are different. We can also see an asymmetry of $\epsilon_2$ in Fig.\ \ref{opticalconstant}. We obtained a better fit (dashed line in Fig.\ \ref{elipsoangle}) to our data with larger TO broadening ($\gamma_{\rm TO}$$>$$\gamma_{\rm LO}$). The mean standard error (difference between model and measured ellipsometric angles, weighted with the experimental errors) is about 10\% smaller with the Lowndes-Gervais model than when fitting with a single Lorentzian. The same observation ($\gamma_{\rm TO}$$>$$\gamma_{\rm LO}$) was made by others\cite{LoYu05} not only for GaP, but also for four other zinc blende semiconductors. Only GaAs is an exception and fulfills the Lorentzian condition $\gamma_{\rm TO}$$\approx$$\gamma_{\rm LO}$. Raman scattering measurements of GaP at room temperature\cite{Ba68} also find $\gamma_{\rm TO}$$>$$\gamma_{\rm LO}$. The TO and LO Raman lineshapes closely follow the peaks of the imaginary part of the dielectric function and the loss function, respectively.\cite{Ba68}

We interpret the GaP phonon broadenings as pure lifetime broadenings. (See supplementary material for a discussion of other potential contributions to the broadenings.) The TO phonon energy at the $\Gamma$-point in the Brillouin zone is at a maximum of the two-phonon density of states. It is almost exactly the same as the sum of the longitudinal and transverse acoustic (LA+TA) phonon energies at the $X$-point. Therefore, the TO phonon can decay very efficiently into two acoustic phonons, while the same is not true for the LO phonon.\cite{VeLo01} The complex self-energy $D\left(\omega_{\rm TO}\right)$ has a large imaginary part. This explains $\gamma_{\rm TO}$$>$$\gamma_{\rm LO}$. 

The phonon dephasing time $\tau$ is related to its broadening by\cite{LaKa78,KuBr84} $\gamma$=$2\hbar/\tau$, where $\hbar$=0.658~meVps is the reduced Planck's constant. The LO phonon lifetime should therefore be larger than that of the TO phonon. This was indeed observed with picosecond pump-probe coherent anti-Stokes Raman scattering (CARS) experiments,\cite{LiKu80} where Kuhl and Bron\cite{KuBr84} found $\tau_{\rm LO}$$\approx$14~ps ($\gamma_{\rm LO}$$\approx$0.76~cm$^{-1}$) and $\tau_{\rm TO}$$<$2.6~ps ($\gamma_{\rm TO}$$>$4 cm$^{-1}$) for GaP at 300~K, see Table \ref{para_Table}. An earlier CARS study\cite{LaLi73} found a longer lifetime of 5.5 ps for a polariton near the TO phonon, resulting in a lower value of $\gamma_{\rm TO}$=2 cm$^{-1}$. 

\subsection{Two-phonon absorption}
\label{two-phonon}

\begin{figure}
\includegraphics[width=\columnwidth]{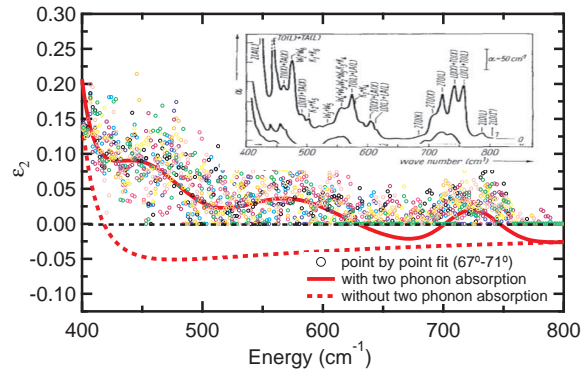}
\caption{Imaginary part of the complex dielectric function for GaP versus photon energy. The inset shows multi-phonon absorption in a high-resolution transmission measurement.\cite{UlJa78}}
\label{multiphonon}
\end{figure}

Berreman and Unterwald\cite{BeUn68} introduced the passivity condition $\gamma_{\rm TO}$$\leq$$\gamma_{\rm LO}$ to ensure that the imaginary part $\epsilon_2\left(\omega\right)$ of the dielectric function does not become negative. This condition is violated for GaP, as described in the previous section. Therefore, our model predicts a negative $\epsilon_2\left(\omega\right)$ just above the LO phonon energy, as shown by the dashed line in Fig.\ \ref{multiphonon}. 

The passivity condition $\epsilon_2$$\geq$0 only applies to the total dielectric function, not to its individual contributions. So far, our Lowndes-Gervais model (\ref{Lowndes}) only includes absorption of light by a single phonon. However, we must also consider multiphonon absorption,\cite{Ba68,KoDa76,HoRu64,KlSp60,UlJa78} where one photon simultaneously creates two, three, or more phonons. 

Multi-phonon absorption is too weak ($\epsilon_2$$<$0.1, $\alpha$$<$100 cm$^{-1}$) to be observable with spectroscopic ellipsometry, which is more appropriate for measuring large absorption coefficients ($\alpha$$>$10$^3$ cm$^{-1}$). Transmission measurements are better suited for the detection of small $\alpha$. We were only able to gain a glimpse of multi-phonon absorption by increasing the band width of our ellipsometer to 4~cm$^{-1}$ and the number of scans fourfold, resulting in the experimental data shown in Fig.\ \ref{multiphonon}. We clearly see that $\epsilon_2$$>$0 in the region from 400 to 600 cm$^{-1}$. We include multiphonon absorption in our model with three broad Gaussians, which are added to Eq.\ (\ref{Lowndes}), as shown by the solid line in Fig.\ \ref{multiphonon}.

These arguments show qualitatively that $\gamma_{\rm TO}$$>$$\gamma_{\rm LO}$ is not unphysical, if the resulting negative values of $\epsilon_2$ are compensated by multi-phonon absorption. The anharmonic decay of optical phonons into acoustic phonons and multi-phonon absorption of a photon are both higher-order processes involving the interaction of several phonons. Therefore, it should not surprise that both processes are related. 

\subsection{Temperature dependence of optical phonon parameters}

To accurately determine the temperature dependence of the optical phonon parameters and the (apparent) surface layer thickness, the ellipsometry angles $\Psi$ and $\Delta$ of bulk GaP at each temperature were fitted with the Lowndes-Gervais model as described in section \ref{methods}. All of these models consisted of a single Lowndes oscillator and the surface layer. The variation of this apparent GaP surface layer thickness with temperature is shown in Fig.\ \ref{Surfacethick} and discussed in the supplementary material. It changes from 0 to 48 {\AA} over the whole temperature range. 

The experimental phonon broadenings are determined by the instrumental resolution and defects (such as polishing damage), which are independent of temperature. In addition, there is a temperature dependence of the broadenings due to the anharmonic decay of optical phonons. In analytical empirical models,\cite{BaKi74,UBaWa83} one usually considers three-phonon processes (where the optical phonon decays into two acoustic phonons of equal energy) and four-phonon processes (where the optical phonon decays into three acoustic phonons of equal energy). {\em Ab initio} calculations\cite{De98} based on realistic phonon dispersions obtained from density-functional theory relax the assumption that all decay products have the same energy, as long as the overall energy and crystal momentum are conserved. Three-phonon decay processes with a 2:1 ratio of the acoustic phonon energies seem particularly important.\cite{MeCa84} The temperature-dependent broadenings can be calculated from the imaginary part of a complex self-energy describing the interaction between the optical phonon and its decay products.\cite{MeCa84} 

The temperature dependence of the phonon energies is given by the corresponding real part of the same self energy.\cite{De00} However, there is also a contribution to the phonon energies due to thermal expansion.\cite{MeCa84} We therefore begin with the discussion of the temperature dependence of the broadenings, followed by the temperature dependence of the phonon energies. We follow the formalism laid out by Bairamov {\em et al.}\cite{BaKi74} and Menendez and Cardona.\cite{MeCa84}

\subsubsection{Temperature dependence of phonon broadenings}

\begin{figure}
\includegraphics[width=\columnwidth]{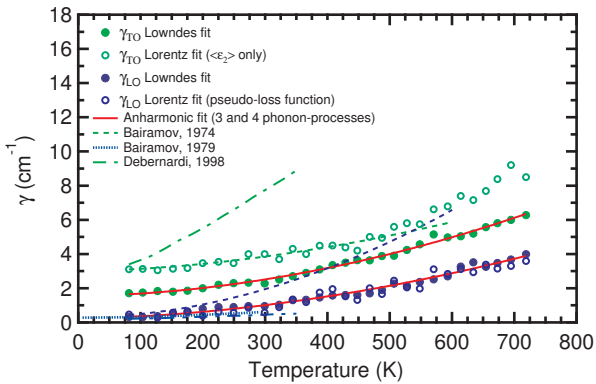}
\caption{Temperature dependence of the TO and LO phonon broadenings (symbols). Solid lines show the best fit to our data using Eq.\ (\ref{BalkanskiBro}) with parameters in Table \ref{BalkanskiBro_Table}. Raman linewidths from Refs.\ \citenum{BaKi74,BaPa79} are shown by dashed and dotted lines, respectively. The dashed-dotted lines show results from density-functional perturbation theory.\cite{De98,De00}}
\label{phononbroadening}
\end{figure}

The broadenings of the TO and LO phonon energies for GaP at each temperature are shown in Fig.\ \ref{phononbroadening}. Solid circles show the results obtained by fitting the ellipsometric angles with a Lowndes model (\ref{Lowndes}). Open circles show results obtained by fitting $\left<\epsilon_2\right>$ and the pseudo-loss function with a Lorentzian. Both methods yield the same values for the LO phonon broadenings, but the TO broadenings found from fitting $\left<\epsilon_2\right>$ are significantly larger than when fitting the ellipsometric angles. This discrepancy is discussed in detail in the supplementary material. 

Our LO broadenings closely follow the 1979 results by Bairamov {\em et al.},\cite{BaPa79} Kuhl and Bron,\cite{KuBr84} and Vall\'ee\cite{Va94} obtained from  Raman linewidths or coherent anti-Stokes Raman scattering decay times of high-purity GaP samples (dotted). Earlier Raman work\cite{BaKi74} used a GaP sample with a higher electron concentration of 9$\times$10$^{16}$ cm$^{-3}$, where the LO broadening is increased due to plasmon-phonon coupling (dashed). The Raman broadenings of the TO phonon\cite{BaKi74} (dashed) are not affected by doping. They follow our results for fitting $\left<\epsilon_2\right>$ with a Lorentzian (open symbols), since the Raman lineshape closely follows $\epsilon_2$, see Ref.\ \citenum{Ba68}. 

Debernardi\cite{De98,De00} calculated the temperature and pressure dependence of the Raman linewidths and energies of the TO and LO phonons from 0 to 325 K, as shown by the dashed-dotted lines. However, his calculated linewidths for the TO phonon agree with Bairamov {\em et al.}\cite{BaKi74} only at low temperatures and are much larger than the experimental results at higher temperatures. In the case of the LO phonon linewidths, the agreement with experiment\cite{BaPa79,KuBr84,Va94} is much better.

Empirically, the broadenings due to anharmonic decay can be fitted with an expression\cite{UBaWa83,SaSa71,BaKi74}
\begin{equation}
\gamma(T)=\gamma_0+A\left[1+\frac{2}{e^x-1}\right]+B\left[1+\frac{3}{e^y-1}+\frac{3}{(e^y-1)^2}\right],
\label{BalkanskiBro}
\end{equation}
where $A$ and $B$ are three- and four-phonon coupling constants, respectively. The reduced phonon energies $x$ and $y$ are
\begin{equation}
x=\frac{\hslash\omega_0}{2k_BT}\quad\text{and}\quad y=\frac{\hslash\omega_0}{3k_BT}.
\label{xy}
\end{equation}
$\omega_0$ is the unrenormalized phonon frequency and $k_B$ is the Boltzmann constant. 
The parameter $\gamma_0$ describes temperature-independent mechanisms, such as inhomogeneous broadening due to defects or the instrumental resolution. Table \ref{BalkanskiBro_Table} lists the best-fit parameters of Eq.\ (\ref{BalkanskiBro}) for TO and LO phonons. 
The broadenings calculated with these parameters (shown by the solid lines in Fig.\ \ref{phononbroadening}) agree very well with the experimental results. Our anharmonic parameters $A$ and $B$ agree well with those in Ref.\ \citenum{BaKi74} for the TO phonon (except for a constant shift related to a different value for $\gamma_0$ due to a different fitting method), but our parameter $B$ is much smaller for the LO phonon, since our sample had a lower doping concentration than in Ref.\ \citenum{BaKi74}. 

\begin{table}[htbp]	
\caption{Anharmonic decay parameters for Eq.\ (\ref{BalkanskiBro}) for the temperature dependence of TO and LO phonon broadenings. Errors are shown in parentheses. Energies marked (f) were fixed during the fit, compare Table \ref{BalkanskiBose_Table}.}
\label{BalkanskiBro_Table}
\begin{ruledtabular}
\begin{tabular}{lll}
This work & TO & LO\\
\hline
$A$ (cm$^{-1}$) & 0.2(1) & 0.2(f) \\
$B$ (cm$^{-1}$) & 0.08(1) & 0.069(2) \\
$\gamma_0$ (cm$^{-1}$) & 1.4(1) &0.03(4) \\
$\omega_0$ (cm$^{-1}$) &367.3(f) & 406.6(f) \\
\hline
Ref.\ \citenum{BaKi74} & TO & LO \\
\hline
$A$ (cm$^{-1}$) & 0.20 & 0.16 \\
$B$ (cm$^{-1}$) & 0.06 & 0.20 \\
$\gamma_0$ (cm$^{-1}$) & 2.8 & 0 \\
$\omega_0$ (cm$^{-1}$) & 366 & 404 \\
\end{tabular}
\end{ruledtabular}
\end{table}

\subsubsection{Temperature dependence of phonon energies}

The temperature dependence of the optical phonon energies due to thermal expansion (TE) is given by\cite{BoMi71,MeCa84,PoFe68}
\begin{equation}
\Omega_{\rm TE}\left(T\right)=
\omega_0\exp\left[-3\gamma\int_0^T\alpha_l\left(\theta\right)d\theta\right],
\label{TE}
\end{equation}
where $\omega_0$ is the phonon energy at 0~K, $\gamma$ the Gr\"uneisen parameter for the phonon mode, and $\alpha_l$ the linear thermal expansion coefficient. Since thermal expansion is small, it suffices\cite{BaKi74} to keep only the linear term in the exponential in Eq.\ (\ref{TE}).

The thermal expansivity can be described using a Debye model for the phonons\cite{RoFa10}
\begin{equation}
\alpha_l\left(T\right)=\alpha_{l\infty}D\left(\frac{\Theta_D}{T}\right),
\end{equation}
where $\Theta_D$=440~K is the Debye temperature, $\alpha_{l\infty}$=5.8$\times$10$^{-6}$ K$^{-1}$ the thermal expansion coefficient in the high temperature limit, and\cite{BaKi74,RoFa10}
\begin{equation}
D\left(\eta\right)=\frac{3}{\eta^3}\int_0^\eta\frac{e^\xi\xi^4}{\left(e^\xi-1\right)^2}d\xi.
\end{equation}
Instead, we can also use tabulated values\cite{De00,DeVo83,La87} for the thermal expansion coefficient and perform the integration in Eq.\ (\ref{TE}) numerically. 

\begin{figure}
\includegraphics[width=\columnwidth]{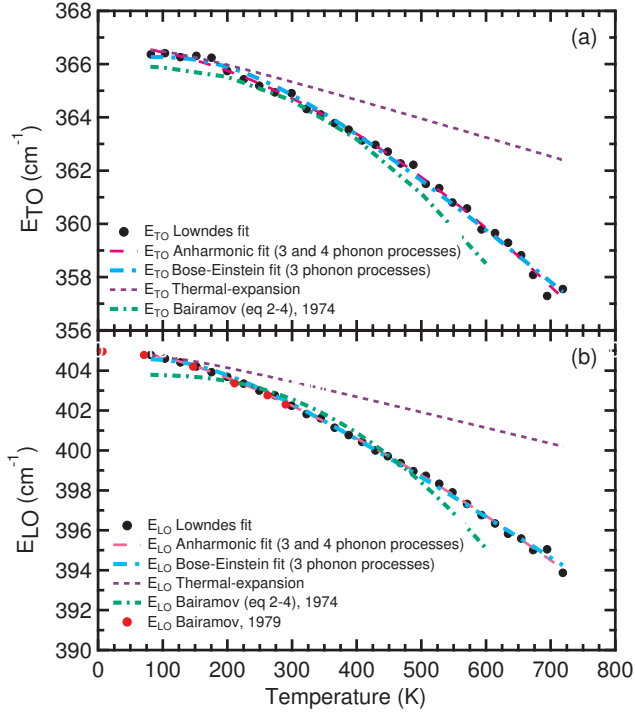}
\caption{Temperature dependence of the (a) transverse (E$_{\rm TO}$) and (b) longitudinal (E$_{\rm LO}$) optical phonon energies of GaP (symbols). The dashed line shows the contribution of thermal expansion. Fits using Eqs.\ (\ref{Balkanski}) and (\ref{Bose}) with parameters in Table \ref{BalkanskiBose_Table} are shown by dashed-dotted lines.}
\label{phononenergy}
\end{figure}

The dependence of TO and LO phonon energies on temperature determined from a Lowndes fit to the ellipsometric angles is shown in Fig.\ \ref{phononenergy} (symbols). Both show a redshift (decrease) with increasing temperature. The contribution of thermal expansion to these redshifts (dashed) is not sufficient to explain the experimental data. Therefore, Bairamov {\em et al.}\cite{BaKi74} also included three- and four-phonon anharmonic decay processes. 

The anharmonic contribution to the redshift can be described as\cite{UBaWa83}
\begin{equation}
\Omega(T)=\omega_0-C\left[1+\frac{2}{e^x-1}\right]-D\left[1+\frac{3}{e^y-1}+\frac{3}{(e^y-1)^2}\right],
\label{Balkanski}
\end{equation}
where and $C$ and $D$ are the anharmonic coupling strengths. 
The second and third terms in Eq.\ (\ref{Balkanski}) relate to three- and four-phonon decay processes, respectively. Setting $D$=0 (i.e., ignoring four-phonon processes) does not achieve a good fit. 

In principle, one should subtract the redshift due thermal expansion from the observed redshift and then fit the parameters $C$ and $D$. This was done by Bairamov {\em et al.}\cite{BaKi74}, which resulted in the parameters shown in Table \ref{BalkanskiBose_Table}. In practice, however, one often ignores the thermal expansion contribution\cite{UBaWa83} and directly fits the experimental data with Eq.\ (\ref{Balkanski}). These parameters are also shown in Table \ref{BalkanskiBose_Table}. The two methods result not only in different sets of coupling constants $C$ and $D$, but also in differences for the unrenormalized phonon energies $\omega_0$. 

This redshift can also be fitted empirically with a Bose-Einstein expression\cite{ViLo84}
\begin{equation}
\Omega(T)=\omega_0-C\left[1+\frac{2}{e^\frac{\theta_B}{T}-1}\right]
\label{Bose}
\end{equation}
as an anharmonic three-phonon decay with an adjustable effective phonon energy $\omega_{\rm eff}$=$k_B\theta_B$, where $\theta_B$ is an effective phonon temperature related to the Debye temperature of GaP (440 K). These results are also shown in Table \ref{BalkanskiBose_Table}.

\begin{table}[htbp]	
\caption{Anharmonic decay parameters for the temperature dependence of TO and LO phonon energies. Errors are shown in parentheses.}
\label{BalkanskiBose_Table}
\begin{ruledtabular}
\begin{tabular}{lll}
Anharmonic decay parameters & TO & LO \\
\hline
\multicolumn{3}{l}{This work, Eq.\ (\ref{Balkanski}), ignoring thermal expansion.} \\
\hline
$\omega_0$ (cm$^{-1}$)  &367.3(2)& 406.6(1) \\
$C$  (cm$^{-1}$) & 0.5(1) & 1.6(1) \\
$D$ (cm$^{-1}$) & 0.14(1) & 0.11(1) \\
\hline
\multicolumn{3}{l}{Ref.\ \citenum{BaKi74}, Eq.\ (\ref{Balkanski}), Raman, thermal expansion subtracted.} \\
\hline
$\omega_0$ (cm$^{-1}$)  & 364.5 & 401.4  \\
$C$  (cm$^{-1}$) & $-$1.76 & $-$2.98 \\
$D$ (cm$^{-1}$) & 0.30 & 0.51 \\
\hline
\multicolumn{3}{l}{This work, Eq.\ (\ref{Bose}), ignoring thermal expansion.} \\
\hline
$\omega_0$ (cm$^{-1}$) &374.6(8)& 410.5(4) \\
$C$ (cm$^{-1}$) & 8.3(9) & 5.9(4) \\
$\omega_{\rm eff}$ (cm$^{-1}$) & 530(30) & 380(20) \\
\end{tabular}
\end{ruledtabular}
\end{table}

\subsubsection{Temperature dependence of high-frequency and static dielectric constants}

\begin{figure}
\includegraphics[width=\columnwidth]{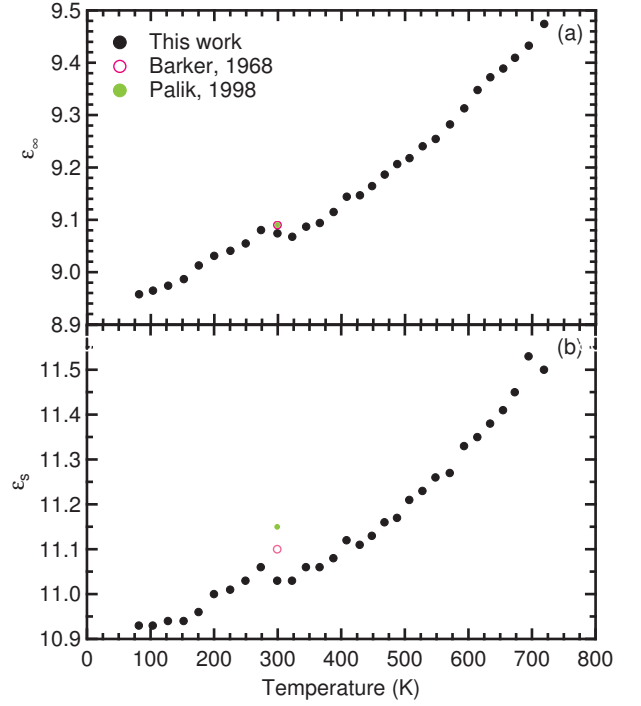}
\caption{Temperature dependence of the (a) high-frequency ($\epsilon$$_{\infty}$) and (b) static ($\epsilon_s$) dielectric constants. $\epsilon_s$ was calculated from the Lyddane-Sachs-Teller relation (\ref{LST}). Data from the literature\cite{Ba68,BoGu98,PaCo71} are also shown.}
\label{epssepinf}
\end{figure}

The high frequency dielectric constant $\epsilon_\infty$ for bulk GaP was also obtained from fitting the ellipsometric angles at each temperature with the Lowndes model (\ref{Lowndes}). It increases with temperature, as shown in Fig.\ \ref{epssepinf}. $\epsilon_\infty$ can be expressed as \cite{YuCa10}
\begin{equation}
\epsilon_{\infty}=1+\left(\frac{E_u}{E_{\rm Penn}} \right)^2, 
\label{epinfinity}
\end{equation}
where the unscreened plasma frequency of the valence electrons is given by
\begin{equation}
E_u^2=\hbar^2\omega_u^2=\frac{\hslash^2Ne^2}{m_0\epsilon_0}=\left({\rm 16.5~eV}\right)^2.
\label{EPenn}
\end{equation}
$N$ is the density of valence electrons per unit volume (8 electrons per formula unit), $e$ the charge of the electron, $m_0$ the free electron mass, and $\epsilon_0$  the vacuum permittivity. The resulting Penn gap $E_{\rm Penn}$=5.83 eV, calculated from $\epsilon_\infty$ at 300~K, is the average separation between the valence and conduction band across the Brillouin zone, usually located near the $E_2$ gap (5.28 eV for GaP, see Refs.\ \citenum{ZoGa93,YuCa10}). We assume that $E_2$ and $E_{\rm Penn}$ have the same temperature dependence.

Taking the derivative of Eq.\ (\ref{epinfinity}) with respect to temperature yields\cite{VeKi75,NeSp12}
\begin{equation}
\frac{d\epsilon_\infty}{dT}=
-3\alpha_l\left(\epsilon_\infty-1\right)-2\left(\epsilon_\infty-1\right)\frac{d\ln{}E_2}{dT}.
\label{derivative}
\end{equation}
The first term considers the decrease of the electron density due to thermal expansion, which results in a decrease of $\epsilon_\infty$. The second term describes the increase of $\epsilon_\infty$ due to the decrease of the Penn gap with increasing temperature. This is the dominant term. 

We calculated the temperature dependence of $\epsilon_\infty$ at room temperature, resulting in 
\begin{equation}
\left.\frac{d\epsilon_\infty}{dT}\right|_{\rm 300~K}=\left(-1.2+7.6\right)\times10^{-4}~{\rm K}^{-1}=6.4\times10^{-4}~{\rm K}^{-1}.
\label{numeric}
\end{equation}
This value compares favorably with the experimental variation of $\epsilon_{\infty}$ between 200 and 400 K, which is 4.5$\times$10$^{-4}$~K$^{-1}$. Therefore, we conclude that the model (\ref{derivative}) for the temperature dependence of $\epsilon_\infty$ is reasonable.

The temperature dependence of $\epsilon_s$ was calculated from the Lyddane-Sachs-Teller (LST) relation (\ref{LST}) and is also shown in Fig.\ \ref{epssepinf}. Since the ratio of the TO and LO phonon energies does not strongly depend on temperature, the variations of $\epsilon_s$ with temperature are very similar to those of $\epsilon_\infty$. We find that $\epsilon_s$ changes by about 1\% between 0 and 300~K, which is three times larger than the change calculated by Debernardi.\cite{De00}

\section{Summary}
We used FTIR spectroscopic ellipsometry from 0.03-0.60 eV to determine the effects of temperature on the properties of long-wavelength optical phonons of GaP at temperature ranging from 80 to 720 K. With different experimental conditions, we also observed two-phonon absorption. GaP shows an asymmetric reststrahlen line shape, because the linewidth of the TO phonon is larger than that of the LO phonon at all temperatures. This violates the Lowndes passivity condition and the imaginary part $\epsilon_2$ of the dielectric function becomes slightly negative just above the LO energy. To achieve a physically meaningful non-negative total dielectric function, we must also add multi-phonon absorption to the single phonon absorption. We also observed that the energies of the TO and LO optical phonons show a redshift and increasing broadening with increasing temperature due to thermal expansion and anharmonic phonon-phonon decay. The static and high-frequency dielectric constant increased with increasing temperature because of the thermal expansion of the crystal and the temperature dependence of the Penn gap. 

\section*{ACKNOWLEDGEMENTS}
This work was supported by the National Science Foundation (DMR-1505172) and by the US Army (W911NF-16-1-0492). We wish to thank Professor David J.\ Lockwood (National Research Council of Canada) and Professor Sotirios Ves (Aristotle University of Thessaloniki, Greece) for helpful discussions regarding this research topic.

\section*{DATA AVAILABILITY}
The data that support the findings of this study are available from the corresponding author upon reasonable request.

\section*{Supplementary Material}

See supplementary material at [URL will be inserted by Publisher] for additional information regarding analytical models for infrared lattice absorption, frequency-dependent broadening, accurate determination of phonon broadening parameters from ellipsometry data, origins of the phonon broadenings, GaP phonon dispersion, the Born effective charge, and the variation of the apparent GaP surface layer thickness with temperature.


\makeatletter
\def\@currentlabel{S1}
\label{something}
\makeatother

\clearpage

\setcounter{page}{1}
\renewcommand{\thepage}{S\arabic{page}}
\setcounter{figure}{0}
\renewcommand\thefigure{S\arabic{figure}}
\setcounter{equation}{0}
\renewcommand\theequation{S\arabic{equation}}
\setcounter{section}{0}
\renewcommand\thesection{S\arabic{section}}
\setcounter{table}{0}
\renewcommand\thetable{S\Roman{table}}

\title{Supplementary Material: \mytitle} 

\makeatletter
{%
  \title@column\titleblock@produce
\makeatother

\noindent
Nuwanjula S.\ Samarasingha and Stefan Zollner

{\small \it \noindent
Department of Physics, New Mexico State University, P.O.\ Box 30001, Las Cruces, NM 88003, USA \\}


\section{Lorentz model for a single phonon}

The static and high-frequency dielectric constants $\epsilon_s$ and $\epsilon_\infty$ and the optical phonon parameters (amplitude $A$, TO and LO phonon energies $\omega_{\rm TO}$ and $\omega_{\rm LO}$, and corresponding broadenings $\gamma_{\rm TO}$ and $\gamma_{\rm LO}$) were obtained by fitting the ellipsometric angles of GaP at each temperature either with a single Lorentzian or with the Lowndes-Gervais model.\cite{Lo70,GePi74} The difference between these two models is that the Lowndes-Gervais model assigns two different broadening parameters $\gamma_{\rm TO}$ and $\gamma_{\rm LO}$ to the two phonons, while the Lorentz model has only one broadening parameter $\gamma_{\rm TO}$.

Following Helmholtz (Ann.\ Phys.\ {\bf 230}, 582, 1875) or Wooten ({\em Optical Properties of Solids}, Academic, New York, 1972), the dispersion due to damped vibrations of molecules in a solid under the influence of an electromagnetic wave can be described by a Lorentz oscillator\cite{Ba68}
\begin{equation}
\epsilon\left(\omega\right)=\epsilon_{\infty}+\frac{A\omega_{\rm TO}^{2}}{\omega_{\rm TO}^{2}-\omega^{2}-i\gamma_{\rm TO}\omega}.
\label{Lorentz}
\end{equation}
TO phonons therefore appear as a strong symmetric peak with linewidth $\gamma_{\rm TO}$ (FWHM) just below $\omega_{\rm TO}$ (due to damping) in the imaginary part $\epsilon_2$ of the complex dielectric function.\cite{TrCa15} The maximum height of this peak is 
\begin{equation}
\epsilon_2\left(\omega_{\rm TO}\right)\approx{}A\frac{\omega_{\rm TO}}{\gamma_{\rm TO}}
\end{equation}
for $\gamma_{\rm TO}$$\ll$$\omega_{\rm TO}$.
The peak position $\omega_{\rm TO}$ can usually determined very precisely from the position of the $\epsilon_2$ peak. However, the peak amplitude is influenced by both $A$ and $\gamma_{\rm TO}$. In practice, this leads to parameter correlations when fitting data, especially if the peak shape is not precisely Lorentzian (e.g., due to asymmetry). 

The minimum and maximum of the real part $\epsilon_1$ of the dielectric function are separated by $\gamma_{\rm TO}$. In this region, $\epsilon_1$ shows anomalous dispersion. The LO phonon energy is defined by $\epsilon_1\left(\omega_{\rm LO}\right)$=0. It can be found from the Lyddane-Sachs-Teller (LST) relation\cite{LoYu05}
\begin{equation}
\epsilon_{s}=\epsilon_{\infty}\frac{\omega_{\rm LO}^{2}}{\omega_{\rm TO}^{2}}.
\label{LST}
\end{equation}
The real part $\epsilon_1\left(\omega\right)$ is negative between $\omega_{\rm TO}$ and $\omega_{\rm LO}$. This leads to a region of high reflectivity called reststrahlen band. Representative graphs of the dielectric function, the complex refractive index, and the reflectivity at normal incidence for a Lorentz oscillator are shown by Wooten (1972). The ellipsometric angles at 70$^\circ$ angle of incidence calculated from the Lorentz model are shown by solid lines in Fig.\ \ref{elipsoangle}. It is important to note that the Lorentz model shows strong symmetry in the dielectric function, the reflectivity at normal incidence, and in the ellipsometric angles (but not in the complex refractive index or the absorption coefficient). For $\gamma_{\rm TO}$=0, the reflectance $R$ equals unity and the ellipsometric angle $\psi$=45$^\circ$ within the reststrahlen band. The corners of $R$ and $\psi$ are very sharp. The rise of $\psi$ or $R$ occurs precisely at the TO frequency, but the drop is pushed beyond the LO frequency at higher angles of incidence.\cite{Sc04} These corners are rounded symmetrically once $\gamma_{\rm TO}$ increases and $R$ and $\psi$ are lower than their ideal undamped values. 
The most precise determination of the broadening parameter $\gamma_{\rm TO}$ can therefore be made from the ellipsometric angle $\psi$ within the reststrahlen region and its deviation from 45$^\circ$. This is independent of the value of the amplitude $A$.
Within the reststrahlen band, there is enhanced sensitivity to the low absorption caused by two-phonon processes (J.\ Huml\'\i\v{c}ek, Thin Solid Films {\bf 313-314}, 687, 1998). 
Setting $\omega$$=$$0$ in Eq.\ (\ref{Lorentz}), we find the relationship
\begin{equation}
\epsilon_s=\epsilon_\infty+A
\label{sinfinity}
\end{equation}
between the static ($\omega$$=$$0$) and high-frequency ($\omega$$\rightarrow$$\infty$) dielectric constants. 
Combining Eqs.\ (\ref{LST}) and (\ref{sinfinity}), we find 
\begin{equation}
\omega_{\rm LO}=\omega_{\rm TO}\sqrt{1+\frac{A}{\epsilon_\infty}}.
\label{LOTO}
\end{equation}

As shown in Fig.\ \ref{loss}, the pseudo-loss function
\begin{equation}
{\rm Im}\left(-\frac{1}{\left<\epsilon\right>}\right)=-\left<\eta_2\right>=\frac{\left<\epsilon_{2}\right>}{\left<\epsilon_{1}\right>^{2}+\left<\epsilon_{2}\right>^{2}} 
\label{Loss}
\end{equation}
shows a strong peak at the LO phonon energy (Wooten, 1972). To obtain the amplitude $B$, energy $\omega_{\rm LO}$, and broadening $\gamma_{\rm LO}$ of the LO phonon, the experimental data points in the pseudo-loss function (\ref{Loss}) were fitted in IGOR Pro with the imaginary part of a Lorentzian
\begin{equation}
-\left<\eta_2\right>=\frac{B\omega_{\rm LO}^2\gamma_{\rm LO}\omega}{({\omega_{\rm LO}^2-\omega^2})^2+\gamma_{\rm LO}^2\omega^2}.
\label{eta}
\end{equation}
The parameters from this fit are shown in model (3) of Table \ref{para_Table}. The LO phonon energy of 402.3 cm$^{-1}$ obtained from this pseudo-loss function fit matched perfectly with the calculated value 402.33 cm$^{-1}$ from Eq.\ (\ref{LOTO}). 
                             
\begin{figure}
\includegraphics[width=\columnwidth]{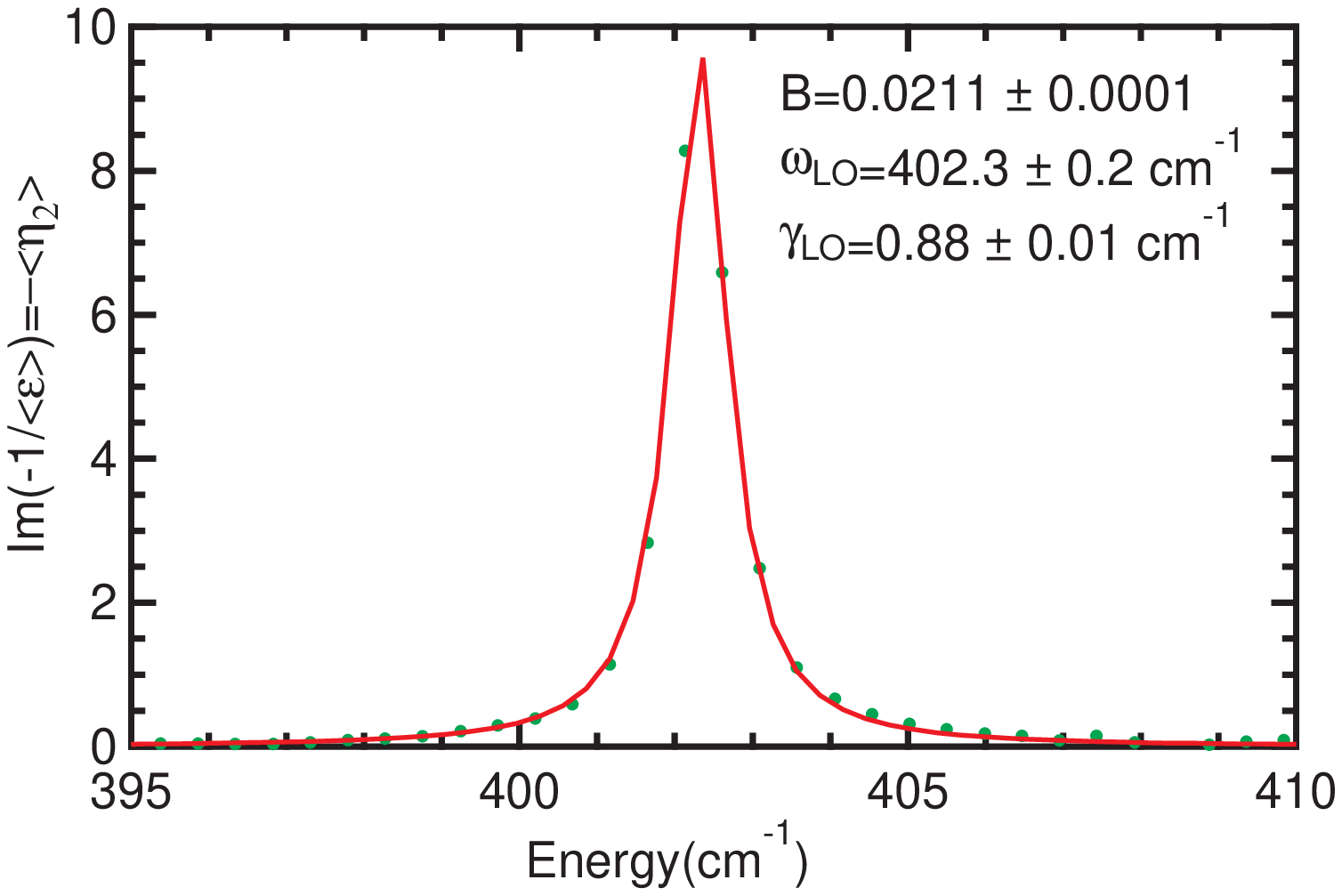}
\caption{Pseudo-loss function for GaP in the reststrahlen region at room temperature (in air) fitted with a Lorentzian as in Eq.\ (\ref{eta}).}
\label{loss}
\end{figure}

\section{Frequency-dependent broadening}

The Lorentz oscillator function (\ref{Lorentz}) can be solved for the broadening parameter
\begin{equation}
i\gamma\omega=
\omega_{\rm TO}^2\frac{\epsilon-\epsilon_s}{\epsilon-\epsilon_\infty}-\omega^2.
\label{gammaTO}
\end{equation}
If $\epsilon$ follows a Lorentzian lineshape, then of course this expression (\ref{gammaTO}) yields a constant $\gamma_{\rm TO}$. However, we can also evaluate this expression for other types of lineshapes or from experimental data $\epsilon\left(\omega\right)=\epsilon_1+i\epsilon_2$, if the static and high-frequency dielectric constants $\epsilon_s$ and $\epsilon_\infty$ are found from extrapolation. 

We can break up Eq.\ (\ref{gammaTO}) into real and imaginary components\cite{MaPa91} (see also A.\ Rastogi, K.\ F.\ Pai, T.\ J.\ Parker, and R.\ P.\ Lowndes, in {\em Proceedings of the International Conference on Lattice Dynamics, Paris, September 5-9, 1977}, edited by M.\ Balkanski (Flammarion, Paris, 1978), p.\ 142)
\begin{eqnarray}
i\gamma\omega&=&\omega_{\rm TO}^2\frac{\left(\epsilon_1-\epsilon_s\right)\left(\epsilon_1-\epsilon_\infty\right)+\epsilon_2^2}{\left(\epsilon_1-\epsilon_\infty\right)^2+\epsilon_2^2}-\omega^2+\nonumber\\
&+i&
\omega_{\rm TO}^2\frac{\left(\epsilon_s-\epsilon_\infty\right)\epsilon_2}{\left(\epsilon_1-\epsilon_\infty\right)^2+\epsilon_2^2}.
\label{gammaTO12}
\end{eqnarray}
In the anharmonic literature,\cite{MaPa91} the broadening term 
\begin{equation}
i\gamma\omega=2\omega_{\rm TO}\left[\Delta\left(\omega\right)+i\Gamma\left(\omega\right)\right]
\end{equation}
in the denominator of the Lorentzian lineshape (\ref{Lorentz})
is typically associated with a complex self-energy $\Delta\left(\omega\right)+i\Gamma\left(\omega\right)$. In this view, $\omega_{\rm TO}$ is the unrenormalized (truly harmonic) resonance frequency of the oscillator. The real part $\Delta\left(\omega\right)$ of this self energy has two contributions due to thermal expansion and due to the anharmonic decay of optical phonons into acoustic ones. It usually leads to a redshift of the phonon frequency. The imaginary part $\Gamma\left(\omega\right)$ causes a broadening of the phonon resonance, leading to a FWHM of $\gamma_{\rm TO}$$\approx$$2\Gamma\left(\omega_{\rm TO}\right)$.

\begin{figure}
\includegraphics[width=\columnwidth]{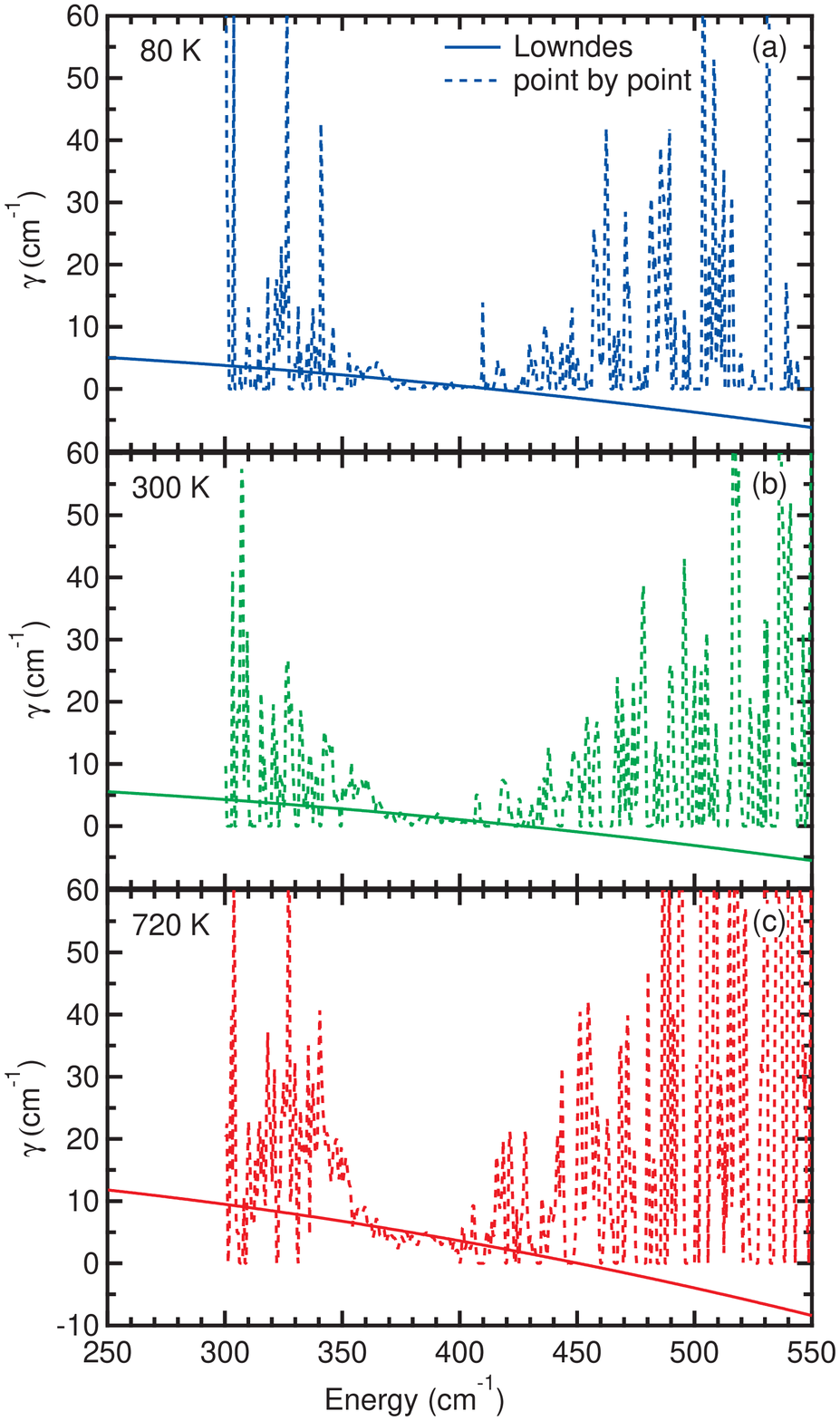}
\caption{Frequency-dependent scattering rate of GaP at three different temperatures calculated from the Lowndes lineshape in Eq.\ (\ref{Lowndes}) (solid) and from our experimental (point-by-point fit) data (dashed) in Fig.\ \ref{opticalconstant}.}
\label{gamma_freq}
\end{figure}

We first evaluate the broadening $\gamma$ in Eq.\ (\ref{gammaTO12}) for a Lowndes lineshape (\ref{Lowndes}), shown by the solid lines in Fig.\ \ref{gamma_freq}. The Lowndes parameters were taken from fits to our ellipsometric angles at three different temperatures. The broadening parameter decreases steadily with increasing photon energy, since $\gamma_{\rm TO}$$>$$\gamma_{\rm LO}$, and eventually becomes negative above the LO frequency where $\epsilon_2$$<$0 for our Lowndes parameters. A negative broadening parameter leads to a pole in the dielectric function below the real axis. This is forbidden, because it violates both the causality and the passivity conditions.\cite{BeUn68} In practice, this is offset by higher-order phonon absorption processes which lead to a positive $\epsilon_2$ as discussed in Sec.\ \ref{two-phonon}. 

The application of Eq.\ (\ref{gammaTO12}) to our experimental data is somewhat disappointing, see the dashed lines in Fig.\ \ref{gamma_freq}. First, we note that $\gamma$ is always positive, because our ellipsometry fitting software enforces $\epsilon_2$$>$0. The noise is very high above the LO phonon energy, where the reflected light intensity is very low. We have excellent signal-to-noise ratio in the reststrahlen region between $\omega_{\rm TO}$ and $\omega_{\rm LO}$, which allows a very accurate determination of the broadening parameter in this region. The data is also noisy below the TO phonon energy, because this spectral region is at the very edge or even below the specified energy range of our instrument. A better signal-to-noise ratio in FTIR ellipsometry would be needed to study the frequency dependent scattering rate.

The broadening parameter $\gamma$ increases with increasing temperature and therefore one expects that this technique should yield better results at higher temperatures, where the broadenings are larger. With a bit of imagination, we might indeed locate a peak in the broadening at 720~K below the TO energy, see Fig.\ \ref{gamma_freq}, where it is expected.\cite{Ba68} See also Ushioda {\em et al.}, Phys.\ Rev.\ B {\bf 8}, 4634 (1973), and Ushioda and McMullen, Solid State Commun.\ {\bf 11}, 299 (1972). Similar results are shown in Ref.\ \citenum{MaPa91} for ZnSe and by Ratogi {\em et al.} (1978) for RbBr and KTaO$_3$. 

\section{Accurate determination of the broadening parameter}

In Fig.\ \ref{phononbroadening}, we show the TO and LO broadenings versus temperature fitted to our experimental data using two different methods: (1) We fitted the ellipsometric angles weighted with the experimental errors with the Lowndes model in Eq.\ (\ref{Lowndes}). (2) We fitted the imaginary part of the pseudodielectric function $\left<\epsilon_2\right>$ 
and the pseudo-loss function $\left<\eta_2\right>$, see Eq.\ (\ref{Loss}), with a Lorentzian to obtain the TO and LO broadenings, respectively. Both methods yield the same broadenings for the LO phonon, but not for the TO phonon. We therefore discuss how the choice of the TO broadenings in the models affects the agreement with the experimental data. Since the Raman lineshapes of the TO and LO phonons closely follow the peaks of $\epsilon_2$ and the loss function,\cite{Ba68} it is not surprising that the second method provides better agreement with the Raman broadenings determined by Bairamov {\em et al.}\cite{BaKi74} 

\begin{figure}
\includegraphics[width=\columnwidth]{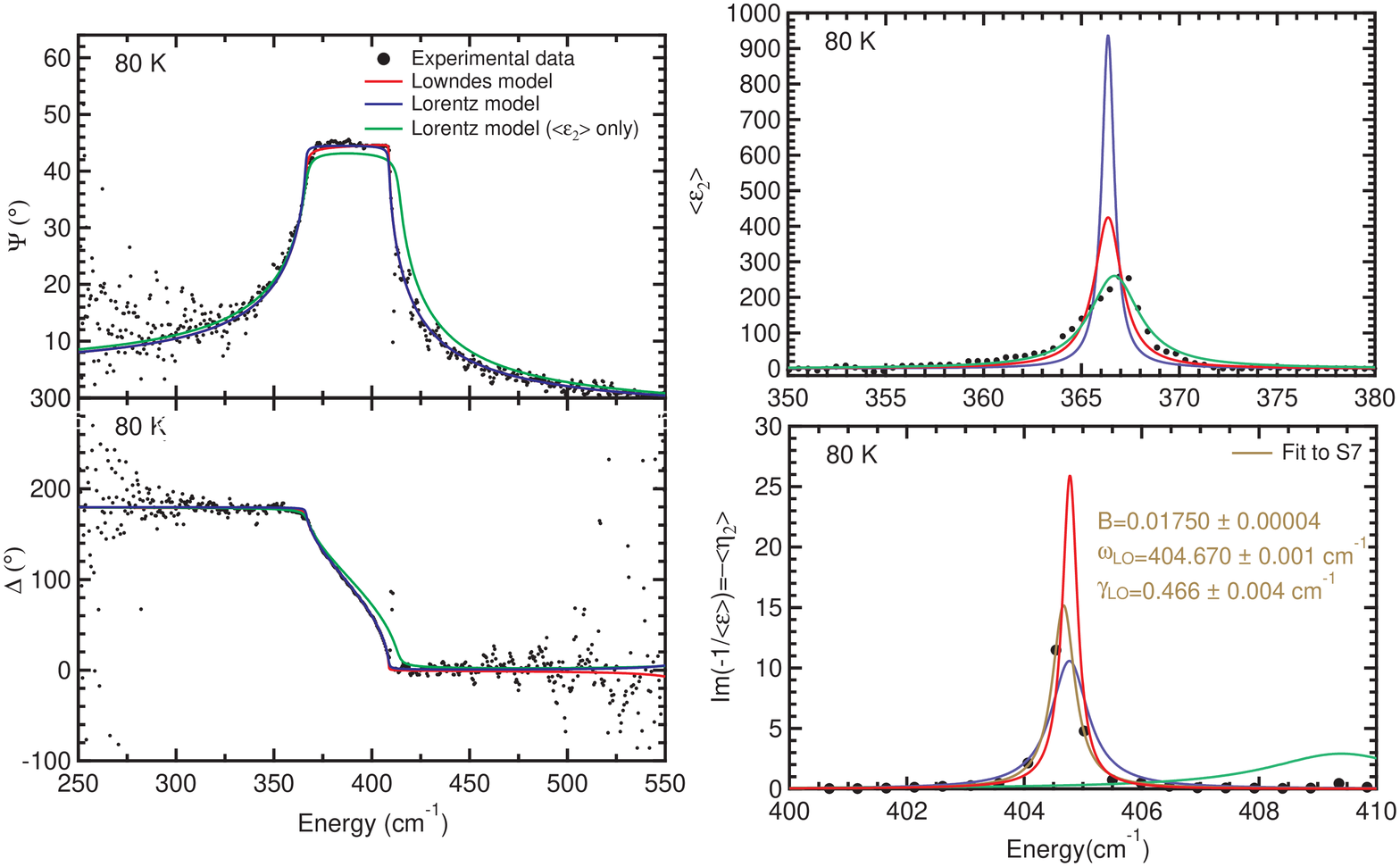}
\caption{Ellipsometric angles $\psi$ and $\Delta$, imaginary part $\left<\epsilon_2\right>$ of the pseudo-dielectric function, and pseudo-loss function $-\left<\eta_2\right>$ versus photon energy for GaP at 80~K (symbols) fitted with two different Lorentz models (blue, green) with different broadening parameters. A Lowndes fit (red) is also shown. The pseudo-loss function is also fitted with a Lorentzian, compare Eq.\ (\ref{eta}), shown by the brown line.}
\label{PDLoss80}
\end{figure}

\begin{figure}
\includegraphics[width=\columnwidth]{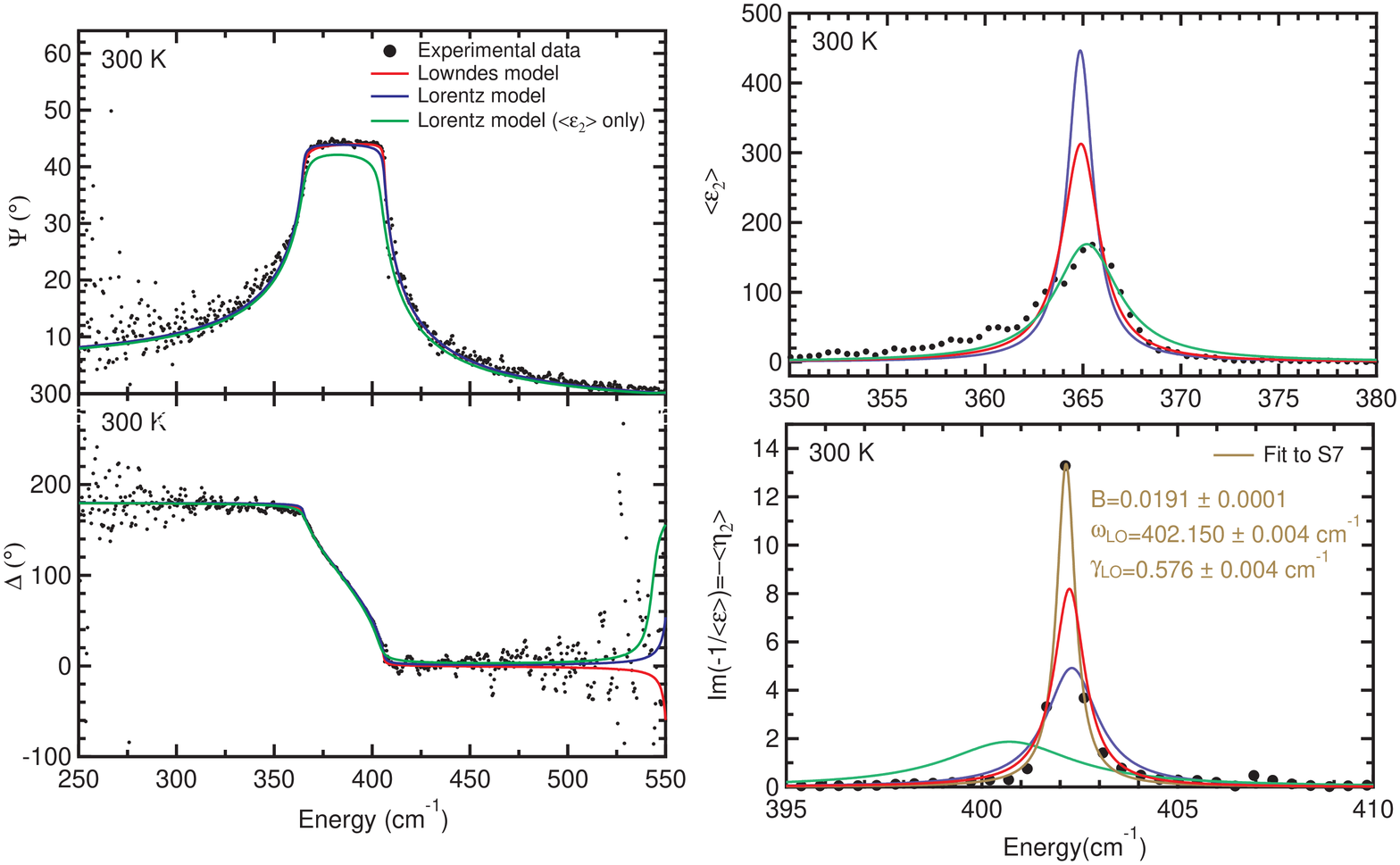}
\caption{Same as Fig.\ \ref{PDLoss80}, but at 300 K (inside the cryostat). The parameters are slightly different from Fig.\ \ref{loss}.} 
\label{PDLoss300}
\end{figure}

\begin{figure}
\includegraphics[width=\columnwidth]{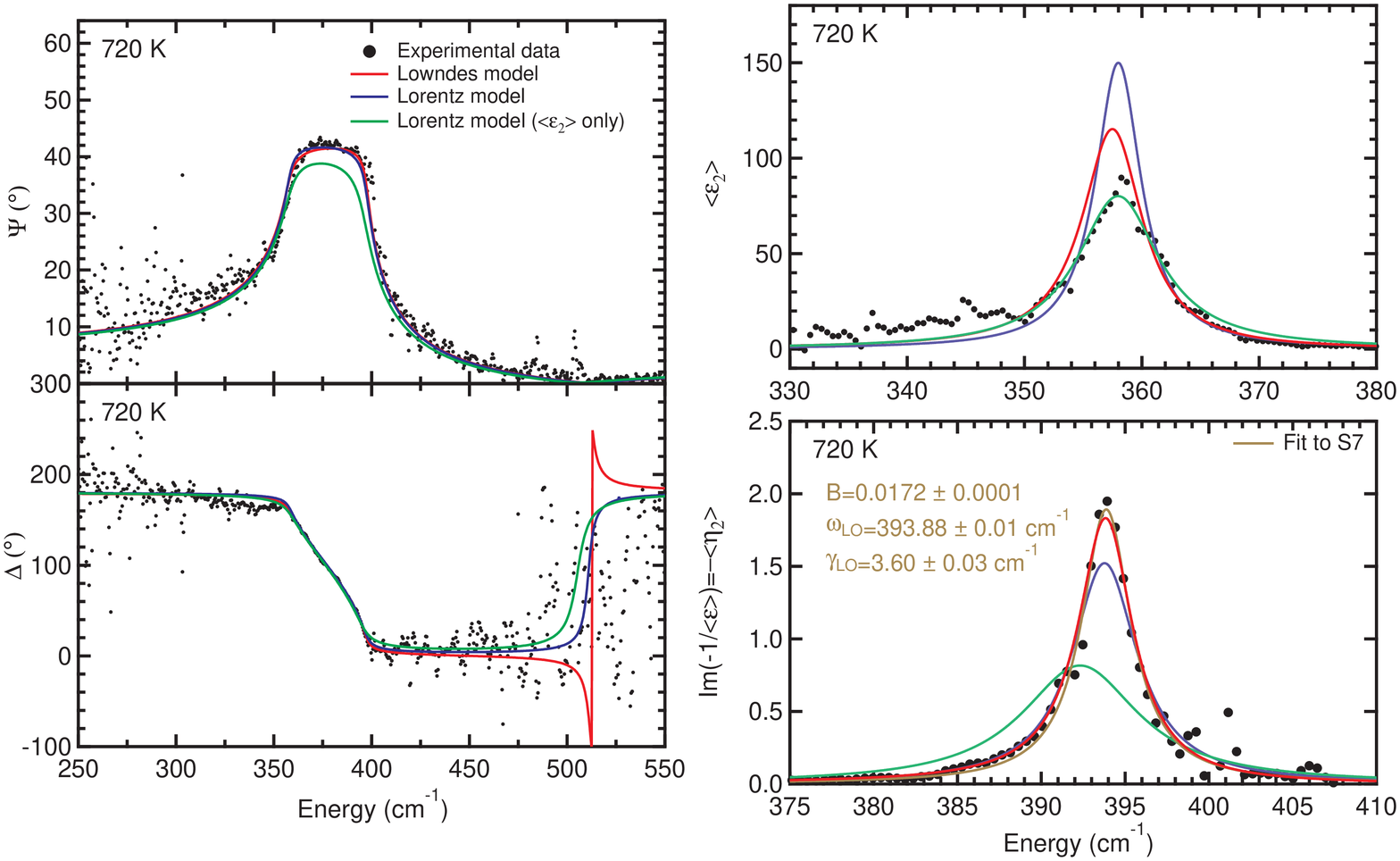}
\caption{Same as Fig.\ \ref{PDLoss80}, but at 720 K.} 
\label{PDLoss720}
\end{figure}

To investigate this question, we show the ellipsometric angles $\psi$ and $\Delta$, the imaginary part $\left<\epsilon_2\right>$ of the pseudo-dielectric function, and the pseudo-loss function $-\left<\eta_2\right>$ versus photon energy for GaP at three different temperatures in Figs.\ \ref{PDLoss80}, \ref{PDLoss300}, and \ref{PDLoss720} (symbols). The Lowndes model (shown by red lines), fitted to the ellipsometric angles, describes the ellipsometric angles and the pseudo-loss function very well, but it significantly overshoots the magnitude of the peak in $\left<\epsilon_2\right>$. Since the resolution of the ellipsometer was set to 1 cm$^{-1}$ and the peak width at room temperature is $\gamma_{\rm TO}$=2.9~cm$^{-1}$, we do not believe that this is an issue with the instrumental resolution. The asymmetry of the $\left<\epsilon_2\right>$ peak also does not explain this discrepancy. Perhaps small systematic errors in the ellipsometric angles lead to large errors in $\left<\epsilon_2\right>$, especially for very large values of $\left<\epsilon_2\right>$. 

Fitting the ellipsometric angles with a Lorentzian (shown in blue in Figs.\ \ref{PDLoss80}, \ref{PDLoss300}, and \ref{PDLoss720}) overshoots the $\left<\epsilon_2\right>$ peak even more. In this case, the fit arrives at a low value of $\gamma$, because the same broadening is used for the TO and LO phonons. The LO phonon broadening is very small and therefore the Lorentz fit takes an average value of the TO and LO broadenings as the Lorentzian broadening $\gamma$, which is too small. Therefore, we understand why the Lorentzian fit to the ellipsometric angles overshoots the $\left<\epsilon_2\right>$ peak. 

Finally, we determine $\gamma_{\rm TO}$ by fitting the $\left<\epsilon_2\right>$ peak with a Lorentzian (green line). We obtain a good fit to the data and the amplitude of the $\left<\epsilon_2\right>$ peak is reproduced well. However, the large broadening leads to a poor fit of the ellipsometric angles and the pseudo-loss function peak. The peak value of the $\psi$-reststrahlen band is much lower in the model than in the data. Also, the falling slope near the LO energy is pushed too far towards larger energies. For completeness, we also show a Lorentzian fit to the pseudo-loss function peak (brown). 

We noted earlier that the peak of the $\psi$-reststrahlen band strongly depends on the broadening. For $\gamma$=0, the maximum of $\psi$ equals 45$^\circ$ and decreases with increasing $\gamma$ between the TO and LO energies. Therefore, accurate measurements of $\psi$ using spectroscopic ellipsometry allow accurate measurements of $\gamma$, even when the instrumental resolution is larger than $\gamma$. While absolute reflectance measurements are very difficult experimentally, the accuracy of $\psi$ is usually better than 0.3$^\circ$ in the mid-infrared spectral range around 30 $\mu$m wavelength. 

It is well known that the magnitude of the reststrahlen bands in GaP reflection measurements is influenced by polishing and etching, presumable due to changes in surface roughness.\cite{KlSp60} A poorly prepared surface will reduce the measured reflectance or the ellipsometric angle $\psi$. Since our value of $\psi$ is above 44$^\circ$ at room temperature, our surface is quality is excellent and does not explain the broad $\left<\epsilon_2\right>$ peak. 

\section{Relationship between Phonon Lifetimes and Broadenings}

According to Laubereau and Kaiser\cite{LaKa78} and Kuhl and Bron,\cite{KuBr84} the relationship between the broadening of the Raman phonon in wave numbers $\Delta\bar\nu$ and the phonon dephasing time (coherence time, or simply lifetime) $\tau$=$T_2$ is 
\begin{equation}
\Delta\bar\nu=\frac{1}{\pi{}c\tau},
\end{equation}
where $c$ is the speed of light. The phonon wave number $\bar\nu$ is related to its wavelength $\lambda$ and its energy $E$ through
\begin{equation}
hc\bar\nu=\frac{hc}{\lambda}=hf=E,
\end{equation}
where $h$ is Plack's constant. 
Therefore, 
\begin{equation}
\gamma=\Delta{}E=hc\Delta\bar\nu=\frac{hc}{\pi{}c\tau}=\frac{h/2\pi}{\pi{}\tau/2\pi}=\frac{\hbar}{\tau/2}=\frac{2\hbar}{\tau}. 
\label{Laubereau}
\end{equation}
Kuhl and Bron\cite{KuBr84} found a dephasing time $T_2$ for GaP at room temperature of 14 ps. The coherent anti-Stokes Raman decay constant $T_2/2$ is therefore 7 ps. The corresponding Raman broadening $\gamma$ equals 0.094 meV or 0.76~cm$^{-1}$, as shown in Fig.\ 2 of Kuhl and Bron.\cite{KuBr84}

A different perspective is given by von der Linde, Kuhl, and Klingenberg.\cite{LiKu80} For GaAs at 77~K, they list a phonon lifetime of $\tau$=6.3~ps and a broadening $\Delta\bar\nu$=0.85 cm$^{-1}$ or $\gamma$=0.10~meV. This suggests a relationship
\begin{equation}
\gamma=\frac{\hbar}{\tau}.
\label{Linde}
\end{equation}
Laubereau, von der Linde, and Kaiser\cite{LaLi73} also emply Eq.\ (\ref{Linde}). 
Equations (\ref{Laubereau}) and (\ref{Linde}) differ by a factor of two. 

In some articles\cite{MaPa91} (R.\ F.\ Wallis, I.\ P.\ Ipatova, and A.\ A.\ Maradudin, Fiz.\ Tverd.\ Tela {\bf 8}, 1064, 1966; Sov. Phys. Solid State {\bf 8}, 850, 1966), the full width at half maximum (FWHM) of a Lorentzian is taken to be $\gamma$=2$\gamma'$. In that case
\begin{equation}
2\gamma'=\frac{\hbar}{\tau} \quad {\rm or} \quad \gamma'=\frac{\hbar}{2\tau}.
\end{equation}

For this work, we follow Eq.\ (\ref{Laubereau}) from Laubereau and Kaiser.\cite{LaKa78}

\section{Origin of the TO and LO Phonon Broadenings}

In the main manuscript we interpreted the TO and LO broadenings as pure lifetime broadenings, but we should also discuss other potential contributions to the phonon broadenings. To this end, we first calculate the optical penetration depth $d_{\rm opt}$ (the inverse of the absorption coefficient $\alpha$) from our model, see Fig.\ \ref{depth}. 

We find that the penetration depth is 200 nm at the TO energy and even larger at different photon energies. Therefore, the width of our phonon peaks is not likely influenced by polishing damage near the surface. Of course, we are not able to rule out other inhomogeneous broadening mechanics, such as crystal defects or impurities. 

\begin{figure}
\includegraphics[width=\columnwidth]{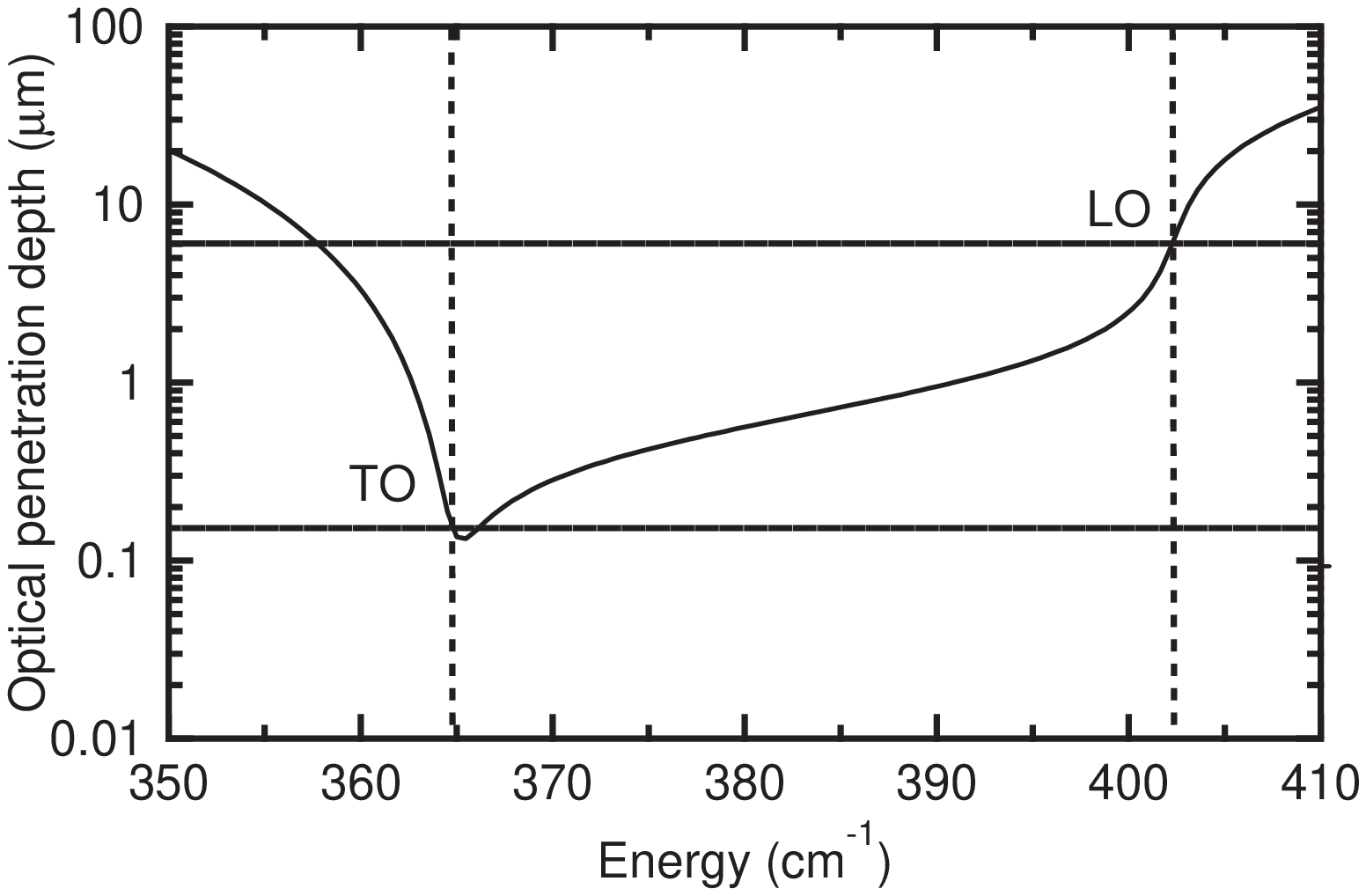}
\caption{Optical penetration depth for GaP in the reststrahlen region as a function of photon energy.}
\label{depth}
\end{figure}

The finite penetration depth also leads to an uncertainty 
\begin{equation}
\Delta{}q=\frac {2\pi}{d_{opt}}\sim0.005\frac {2\pi}{a}
\label{wavevector}
\end{equation}
of the phonon wave vector, where $a$ is the lattice constant of GaP. We find the corresponding uncertainty $\Delta{}E$ by calculating the phonon dispersion of GaP from a ten-parameter shell model\cite{BoHa79} using the programs of Kunc and Nielsen (Comput.\ Phys.\ Commun.\ {\bf 17}, 413, 1979) as shown in Fig.\ \ref{dispersion}. For our (111) oriented sample, the uncertainty is along the $\Lambda$-direction. We find that the finite penetration depth causes an energy broadening of no more than 0.002 cm$^{-1}$, much smaller than our spectrometer resolution (1 cm$^{-1}$). Therefore, we conclude that the finite penetration depth does not have a significant impact on the phonon broadenings of GaP. By contrast, the finite penetration depth of x-rays determines the widths of Bragg reflections in C, Si, and Ge (Pietsch, Holy, Baumbach, {\em High-Resolution X-Ray Scattering}, Springer, New York, 2004, Fig.\ 1.5). 

\begin{figure}
\includegraphics[width=\columnwidth]{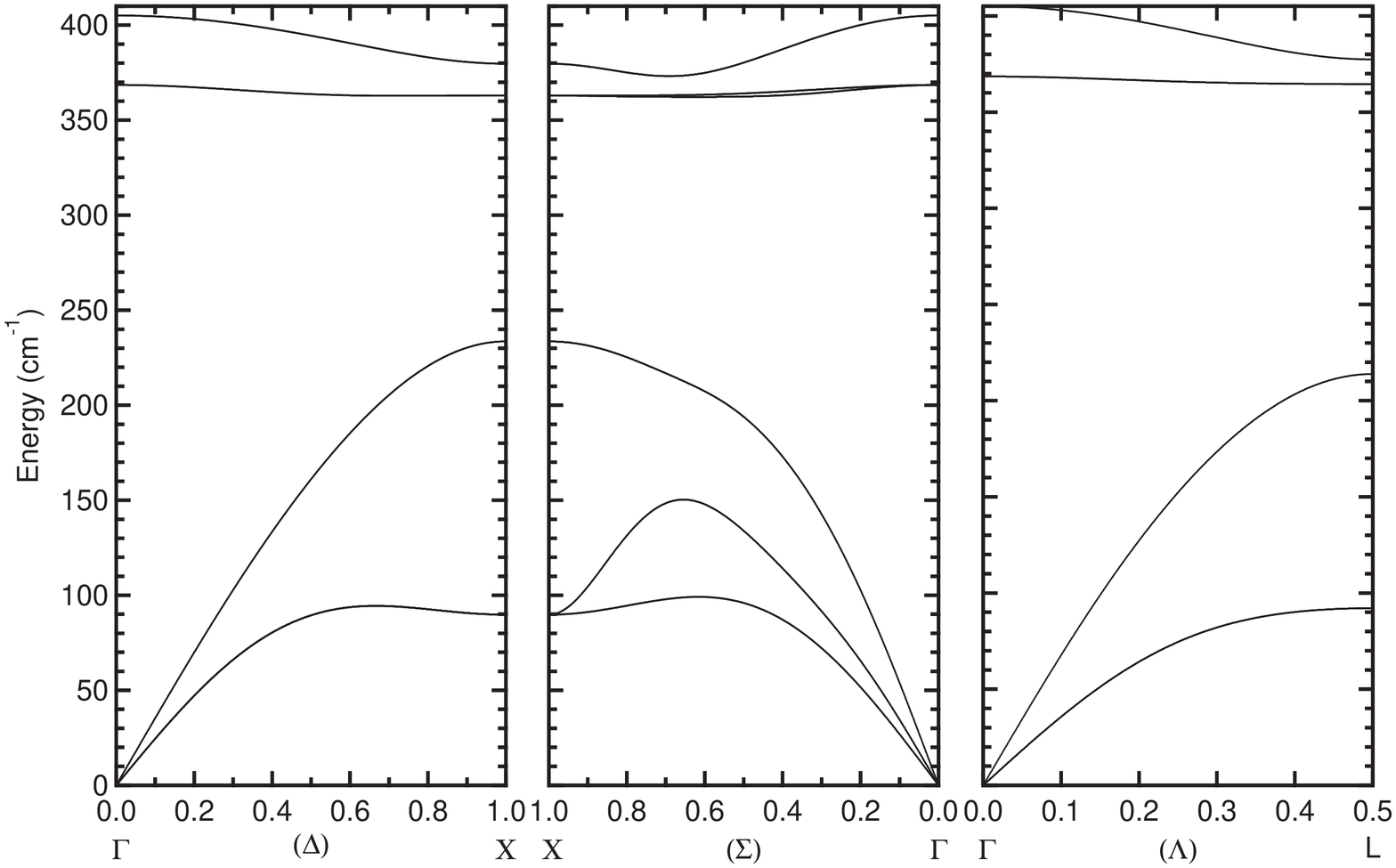}
\caption{Phonon dispersion curves for GaP, calculated using a ten-parameter shell model.\cite{BoHa79}}
\label{dispersion}
\end{figure}

Finally, we discuss the impact of doping on the optical phonon broadenings. For an effective electron Drude (or conductivity) mass of (P.\ K\"uhne, T.\ Hofmann, C.\ M.\ Herzinger, and M.\ Schubert, Thin Solid Films {\bf 519}, 2613, 2011)
\begin{equation}
m_{\rm cc}=\frac{3m_lm_t}{2m_l+m_t}=0.35m_0 
\end{equation}
for GaP in the $X$-valley (where $m_l$=1.12$m_0$ and $m_t$=0.22$m_0$ are the longitudinal and transverse masses, respectively, and $m_0$ is the free electron mass) and an electron concentration near 5$\times$10$^{16}$ cm$^{-3}$, the screened plasma frequency\cite{ZoPa19} $\omega_P$ is about 5 meV or 40 cm$^{-1}$. The lower (LP) and upper phonon-plasmon (UP) resonances are therefore given by (Varga, Phys.\ Rev.\ {\bf 137}, A1896, 1965; Mooradian and Wright, Phys.\ Rev.\ Lett.\ {\bf 16}, 999, 1966; Kukharskii, Solid State Commun.\ {\bf 13}, 1761, 1973)
\begin{equation}
\omega_{\rm LP,UP}^2=
\frac{1}{2}\left(\omega_P^2+\omega_{\rm LO}^2\right)\pm
\sqrt{\frac{1}{4}\left(\omega_P^2+\omega_{\rm LO}^2\right)^2-\omega_P^2\omega_{\rm TO}^2}.
\end{equation}
The UP energy is the experimentally measured LO energy modified by doping. 

Using $\omega_P$=40 cm$^{-1}$, $\omega_{\rm TO}$=364.8 cm$^{-1}$, and $\omega_{\rm LO}$=401.9~cm$^{-1}$, we find $\omega_{\rm LP}$=36.3 cm$^{-1}$ (below our spectral range) and $\omega_{\rm UP}$=402.3 cm$^{-1}$. This means that the ``true'' LO energy for an undoped sample is about 0.4 cm$^{-1}$ below our measured value of 402.3 cm$^{-1}$ for our electron concentration of 5$\times$10$^{16}$~cm$^{-3}$. This might explain small variations of the reported values of $\omega_{\rm LO}$ in the literature, if no doping correction was performed. 

With increasing doping concentration, the UP broadening (i.e., the observed LO broadening due to the interaction between the LO phonon and free carriers) increases considerably, while the TO phonon broadening remains about the same (Kukharskii 1973). 
This was observed in two different studies by Bairamov {\em et al.,}\cite{BaKi74,BaPa79} where the sample with the higher resistivity of 10$^{10}$-10$^{12}$ $\Omega$cm had an LO broadening of 0.6 cm$^{-1}$, while a less pure sample with a carrier concentration of 9$\times$10$^{16}$ cm$^{-3}$ had an LO broadening of 2 cm$^{-1}$, see Table \ref{para_Table}. 
Therefore, free-carrier effects cannot explain why the TO broadening is larger than the LO broadening. We also verified this by adding the effects of free carriers in our ellipsometry model with a Drude term. As shown in Fig.\ \ref{doping}, the ellipsometric angle $\psi$ for a 70$^\circ$ angle of incidence depends on the doping level for carrier concentrations exceeding 5$\times$10$^{17}$ cm$^{-3}$, one order of magnitude higher than for our sample. For this simulation we assumed a mobility of 375 cm$^2$/Vs (independent of carrier concentration), calculated from the resistivity of 0.3 $\Omega$cm specified by the supplier of this sample. 

\begin{figure}
\includegraphics[width=\columnwidth]{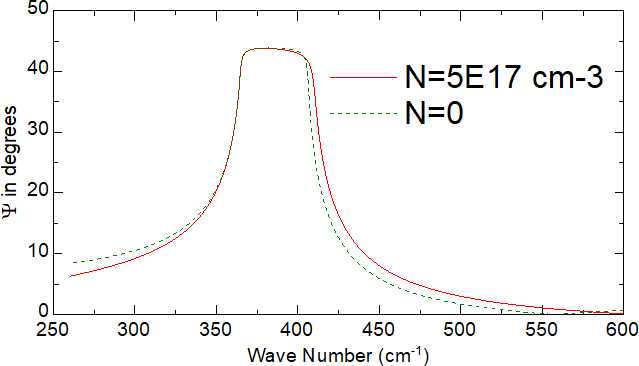}
\caption{Doping dependence of the ellipsometric angle $\psi$ for bulk GaP at room temperature and 70$^\circ$ angle of incidence.}
\label{doping}
\end{figure}

Since we perform measurements at elevated temperatures, one might also ask if the density of thermally excited carriers can contribute to the broadenings of LO phonons. We therefore calculated the intrinsic carrier density (N.\ W.\ Ashcroft and N.\ D.\ Mermin, {\em Solid State Physics}, Saunders, Fort Worth, 1976, p.\ 575)
\begin{eqnarray}
&n_i\left(T\right)=\frac{1}{4}\left(\frac{2k_BT}{\pi\hbar^2}\right)^{3/2}\left(m_cm_v\right)^{3/4}e^{-\frac{E_g}{2k_BT}}=& \label{intrinsic} \\
=&2.5\left(\frac{m_cm_v}{m_0^2}\right)^{3/4}
\left(\frac{T}{300~K}\right)^{3/2}e^{-\frac{E_g}{2k_BT}}\times10^{19}~{\rm cm}^{-3}. & \nonumber
\end{eqnarray}
Here we need to use the density of states mass for electrons
\begin{equation}
m_c=N_v^{2/3}\sqrt[3]{m_lm_t^2}=0.79m_0,
\end{equation}
where $N_v$=3 is the number of $X$-valleys. The other parameters are $m_v$=0.83$m_0$ and $E_g$=2.25 eV. The results of calculating the intrinsic carrier density (\ref{intrinsic}) are shown in Fig.\ \ref{carriers}. Even at our highest temperatures, the intrinsic density of thermally excited carriers is many orders of magnitude lower than the doping concentration of 5$\times$10$^{16}$~cm$^{-3}$ and therefore can be ignored. 

\begin{figure}
\includegraphics[width=\columnwidth]{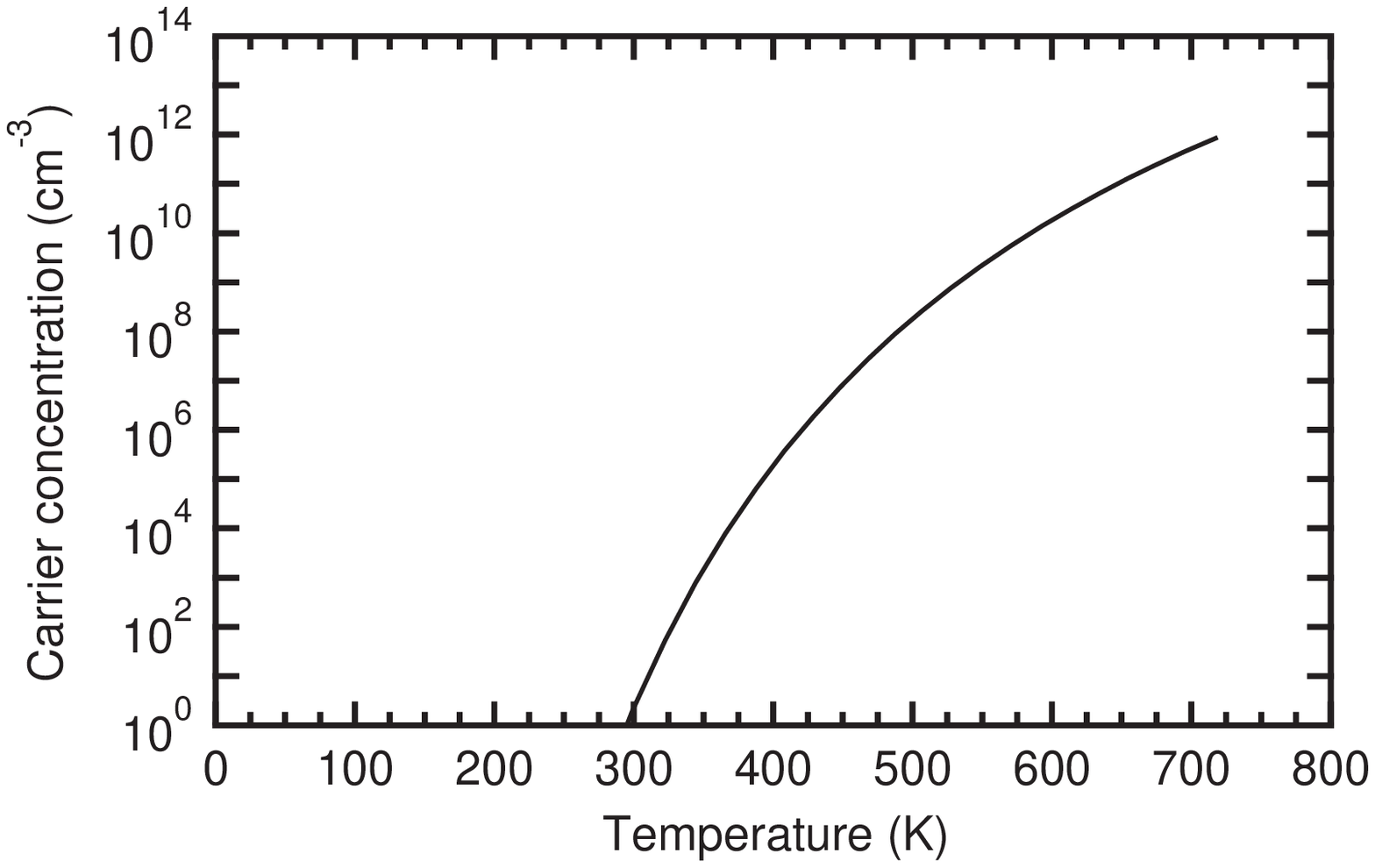}
\caption{Intrinsic carrier concentration of GaP as a function of temperature calculated from Eq.\ (\ref{intrinsic}).}
\label{carriers}
\end{figure}

\begin{figure}
\includegraphics[width=\columnwidth]{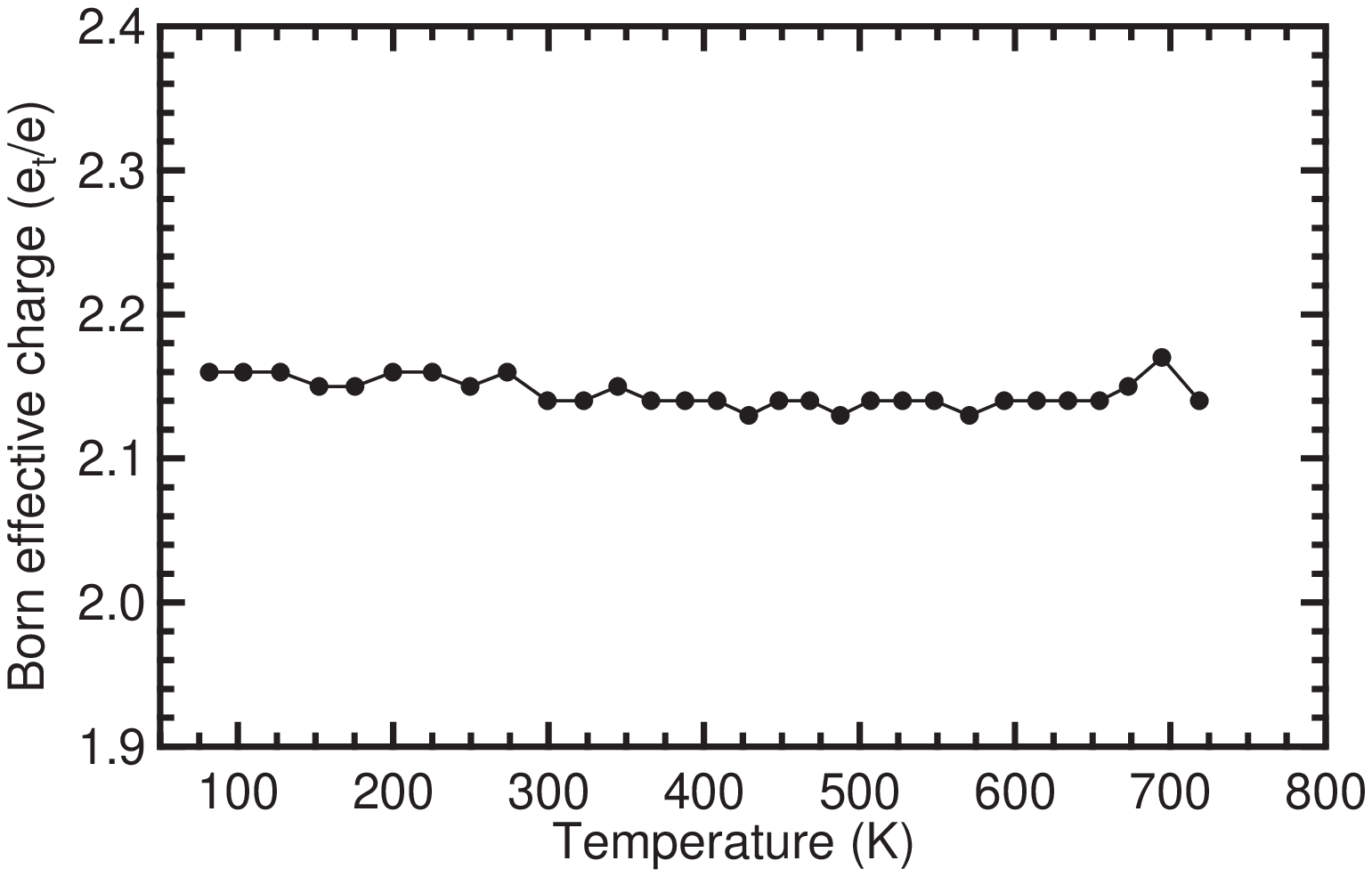}
\caption{Temperature dependence of the Born effective charge calculated from Eq.\ (\ref{BEC}) for bulk GaP.}
\label{BECf}
\end{figure}

\begin{figure}[b]
\includegraphics[width=\columnwidth]{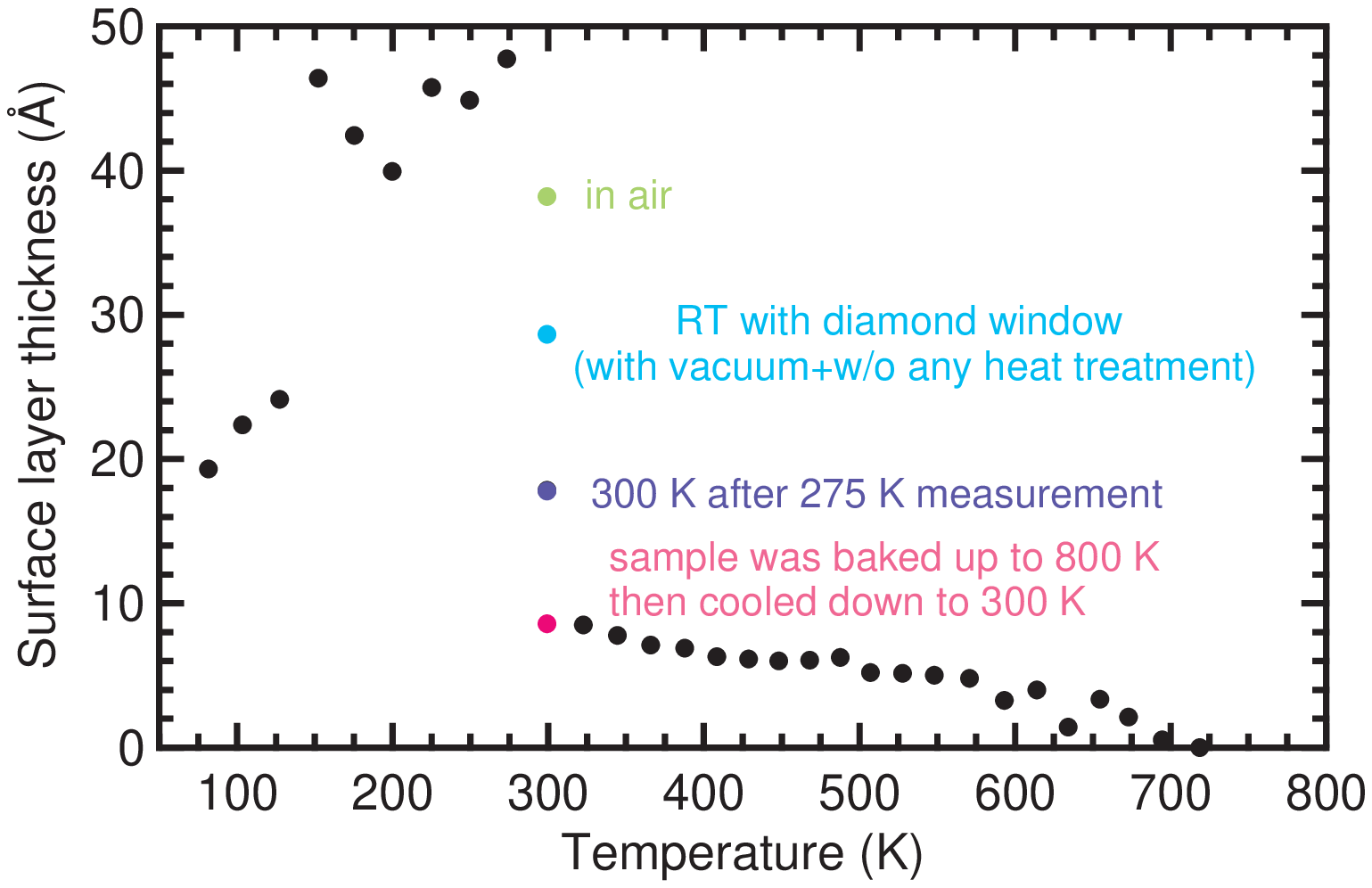}
\caption{Variation of the apparent GaP surface layer thickness with temperature.}
\label{Surfacethick}
\end{figure}

\section{GAP BORN EFFECTIVE CHARGE}

In a polar crystal, there is a charge transfer between the atoms involved in the chemical bond. This is known as the Born effective charge $e_t^*$. It does not have a large influence on the energies of long-wavelength TO phonons, but it adds an additional Coulombic restoring force to the long-wavelength LO vibrations (Fr\"ohlich interaction). This leads to a splitting of the TO and LO phonons at the $\Gamma$-point. From this Fr\"ohlich splitting, the Born effective charge
\begin{equation}
\ e_t^*=\sqrt{V\mu\epsilon_0\epsilon_\infty(\omega_{LO}^2-\omega_{TO}^2 )}
\label{BEC}
\end{equation}
can be calculated at each temperature, see Reparaz {\em et al.}, Appl.\ Phys.\ Lett. {\bf 96}, 231906 (2010). $V$ is the volume per GaP formula unit (volume of the primitive unit cell), $\mu$ the reduced mass of the Ga and P atoms, $\epsilon_0$ the vacuum permittivity, and $\epsilon_\infty$ the high-frequency dielectric constant. As shown in Fig.\ \ref{BECf}, the Born effective charge does not depend on temperature, as one would expect. The average Born effective charge is 2.15(1) over the whole temperature range. We did not observe the small increase in the Born effective charge with temperature due to thermal expansion predicted by Debernardi.\cite{De00}

\section{Variation of the apparent GaP surface layer thickness}

Insulators are mostly transparent in the mid-infrared spectral region, above the phonon absorption energies and below the band gap. In this region, the ellipsometric angle $\Delta$ must be 0 or $\pi$ for a bare substrate, depending on the relative magnitude of the angle of incidence and the Brewster angle. Experimentally, one observes deviations of $\Delta$ from the ideal values for two reasons: (1) Surface overlayers, such as surface roughness, native oxides, or adsorbed overlayers such as water, ice, or other airborne molecular contamination. (2) Systematic errors affecting the measurement of $\Delta$, especially those caused by the diamond windows of our UHV cryostat.\cite{Ab20} 

In Fig.\ \ref{Surfacethick}, we report the surface layer thickness in our experiments, modeled as a 50/50 mixture of GaP and voids within the Bruggeman effective medium approximation. We call this the {\em apparent} surface layer thickness, because it might be affected by systematic errors from the cryostat windows. 

An initial measurement of the GaP sample in air (without windows) finds a surface layer thickness of 38 \AA, shown by the green data point. The sample is then mounted in the cryostat followed by pumping to UHV conditions. This reduces the surface layer thickness to 29 \AA, as shown by the blue data point. This reduction might be due evaporation of some of the overlayer, but it could also be due to errors caused by our windows. We then heat the sample to 800~K for several hours and let it cool down to room temperature overnight. This reduces the surface layer thickness to 9 \AA, as shown by the red data point. Most likely, this value of 9 \AA{} represents the residual surface roughness, after all adsorbed contaminants have evaporated. 

We now start our temperature series at 80~K, where the surface layer thickness is 19 \AA. We gradually increase the temperature to 275~K and measure the pseudo-dielectric function at each step for several hours. This increases the surface layer thickness to 48 \AA, as ice forms on the sample. At the next temperature (300~K), the surface layer thickness is only 18 \AA, because most of the ice has evaporated. As we heat the sample to 325~K, the surface layer thickness is 8 \AA{} and it gradually is reduced to zero as we heat towards 700~K. It is not likely that the surface layer thickness will actually vanish at 700~K. It is more likely that our data are affected by experimental errors due to the diamond windows. These windows are not getting warm, even at the highest sample temperatures. Therefore, the errors due to the windows are likely independent of temperature. It is likely that we underestimate the surface layer thickness in the cryostat by about 10 \AA, due to the retardance of the diamond windows.

This discussion emphasizes that one must be careful when interpreting changes of optical constants with temperature. A careful preparation of the sample is required to minimize surface overlayers. It can be very helpful to heat the sample for several hours to reduce airborne molecular contamination, if this does not change the properties of the sample. It is inevitable for ice to form on the sample below room temperature, even under the best (10$^{-8}$ Torr) vacuum conditions achievable in our setup. 

\clearpage

%
%
%
%
%
%
\end{document}